\documentclass[prb,aps,twocolumn,superscriptaddress,10pt, longbibliography]{revtex4-2}
\usepackage{graphicx}
\usepackage{dcolumn}
\usepackage{bm}
\usepackage{hyperref}
\usepackage[mathlines]{lineno}
\usepackage[utf8]{inputenc}

\usepackage{physics}
\usepackage{chemformula}
\usepackage{simpler-wick}
\usepackage{setspace}
\usepackage[T1]{fontenc}
\usepackage{amsfonts}
\usepackage{amsmath}
\usepackage{xcolor}
\usepackage{enumitem}
\usepackage{makecell}
\usepackage{ulem} 

\hypersetup{
	colorlinks = true,
	allcolors= blue,
}

\renewcommand{\a}{\alpha}
\renewcommand{\b}{\beta}

\newcommand{\w}{\omega}

\renewcommand{\k}{\kappa}
\newcommand{\h}{\hbar}
\newcommand{\p}{^\prime}
\newcommand{\pp}{^{\prime\prime} }

\DeclareMathAlphabet{\mathsfit}{\encodingdefault}{\sfdefault}{m}{sl}
\SetMathAlphabet{\mathsfit}{bold}{\encodingdefault}{\sfdefault}{bx}{sl}
\newcommand{\tens}[1]{\bm{\mathsfit{#1}}}
\newcommand{\tenscomp}[1]{\mathsfit{#1}}

\usepackage[final]{changes}

\makeatletter
\@namedef{Changes@AuthorColor}{magenta}
\colorlet{Changes@Color}{magenta}
\makeatother

\begin{document}

\preprint{APS/123-QED}

\title{Many-body Green's function approach to lattice thermal transport}




\author{Giovanni Caldarelli}
 \email{giovanni.caldarelli@uniroma1.it}
\affiliation{Dipartimento di Fisica, Università di Roma La Sapienza, Piazzale Aldo Moro 5, I-00185 Rome, Italy}

\author{Michele Simoncelli}
\altaffiliation{Present address:  
Theory of Condensed Matter Group of the Cavendish Laboratory and Gonville \& Caius College, University of Cambridge (UK)}
\affiliation{Theory and Simulation of Materials (THEOS) and National Centre for\\
Computational Design and Discovery of Novel Materials (MARVEL),\\
École Polytechnique Fédérale de Lausanne, Lausanne, Switzerland.
}
\author{Nicola Marzari}%
\affiliation{Theory and Simulation of Materials (THEOS) and National Centre for\\
Computational Design and Discovery of Novel Materials (MARVEL),\\
École Polytechnique Fédérale de Lausanne, Lausanne, Switzerland.
}
\author{Francesco Mauri}
\author{Lara Benfatto}
 \email{lara.benfatto@roma1.infn.it }
\affiliation{Dipartimento di Fisica, Università di Roma La Sapienza, Piazzale Aldo Moro 5, I-00185 Rome, Italy}

\date{\today}
\begin{abstract}
Recent progress in understanding thermal transport in complex crystals has highlighted the prominent role of heat conduction mediated by interband tunneling processes, which emerge between overlapping phonon bands (\textit{i.e}.\ with energy differences smaller than their broadenings). These processes have recently been described in different ways, relying on the Wigner or Green-Kubo formalism, leading to apparently different results which question the definition of the heat-current operator. Here, we implement a full quantum approach based on the Kubo formula, elucidating analogies and differences with the recently introduced Wigner or Green-Kubo formulations, and extending the description of thermal transport to the overdamped regime of atomic vibrations, where the phonon quasiparticle picture breaks down. We rely on first-principles calculations on complex crystals with ultralow conductivity to compare numerically the thermal conductivity obtained within the aforementioned approaches, showing that at least in the quasiparticle regime the differences are negligible for practical applications.
\end{abstract}

\maketitle

\section{Introduction}
In the last few years, the technological interest in increasing the efficiency of thermoelectric energy-conversion devices  \cite{wang2018cation,pisoni2014ultra,xia2020microscopic}, or in optimizing thermal shields and thermal barrier coatings  \cite{suresh1997investigation,vassen2000zirconates,chen2009thermophysical,wan2010glass,zhang2020microstructure,yang2016effective}, has stimulated intense research on materials with ultralow thermal conductivity.
So-called complex crystals, defined as materials with a phonon spectrum featuring interband spacings smaller than the linewidths  \cite{simoncelli2019unified,long_paper}, are promising candidates for these applications since their thermal conductivity is very low (typically ${\lesssim} 1 \tfrac{\rm W}{m\cdot K}$ around room temperature).
More precisely, the thermal properties of complex crystals can be regarded as intermediate between those of simple crystals, where the interband spacings between phonon branches are much larger than their linewidths  \cite{simoncelli2019unified,long_paper}, and those of glasses, where vibrational eigenstates are quasi-degenerate. More specifically, while in simple crystals the thermal conductivity $\kappa(T)$ follows the typical \cite{ziman2001electrons} Peierls-Boltzmann $\kappa(T){\sim} T^{-1}$ decay for $T{\gg} \theta_D$ (where $\theta_D$ is the Debye temperature), in complex crystals $\kappa(T)$  has a much milder asymptotic decay, which resembles the saturating trend typical of glasses  \cite{allen1989thermal,feldman1993thermal,Feldman1995,allen1999diffusons}.  
Such intermediate behavior has been related by several works  \cite{donadio2009atomistic,chen2015twisting,mukhopadhyay2018two,isaeva2019modeling,simoncelli2019unified,luo2020vibrational} to the coexistence of Peierls-Boltzmann intraband transport, dominant in simple crystals \cite{hardy1963energy,peierls1929kinetischen,peierls1955quantum}, and interband transport, dominant in glasses  \cite{allen1989thermal} \footnote{Here the term ``interband'' for glasses is understood considering glasses as limiting cases of disordered but periodic crystals in the limit of infinitely large primitive cells}.

 The milestone work by Hardy \cite{hardy1963energy} that generalized the heat-flux operator
	in a lattice for interband transport stimulated some theoretical work in the 1960s aiming
	at a formal description of both mechanisms within a Green-Kubo formalism\cite{maradudin1964lattice,semwal1972thermal}, but
	a quantitative comparison among the two mechanisms emerged only in recent times. The
	reason for such delay lies probably in the fact that the computing power needed to simulate
	systems in which interband transport is relevant was prohibitive for the time. In fact, it
	was not long ago that very accurate first-principle scattering rates to be used in thermal
	transport calculation became available  \cite{narasimhan1991anharmonic,debernardi1995anharmonic,broido2007intrinsic}. Thus, a number of works have 
	relied on these advances to study heat conduction from first-principles in crystals using the simplest numerically manageable model, the Peierls-Boltzmann equation for intraband transport, highlighting its accuracy  in simple crystals  \cite{broido2007intrinsic,PhysRevLett.106.045901,Esfarjani_PRB_11,Chen_Science_12,Nanoscale_Thermal_14,mcgaughey2019phonon,paulatto2015first,fugallo2013ab,fugallo2014thermal,cepellotti2015phonon,cepellotti2016thermal} but also its failures in complex crystals \cite{donadio2009atomistic,li2015ultralow,lee2017ultralow,chen2015twisting,Weathers_PRB_2017,lory2017direct,mukhopadhyay2018two}.
	As a result, these works have sparked interest in understanding how to describe accurately --- and in a computationally affordable form --- thermal transport.
	The most recent theoretical efforts  \cite{pereverzev2018theoretical,simoncelli2019unified,isaeva2019modeling,dangic2020origin} were thus motivated by a quantitative estimate of interband effects, but they actually led to a broader perspective about the microscopic description of heat transport, highlighting the failures of Peierls-Boltzmann formulation in complex crystals  \cite{simoncelli2019unified}, where interband transitions emerge.

The existence of different approaches to account for interband transport raises the question of elucidating analogies and differences between them, and most importantly, benchmarking the differences in their predictions for the thermal conductivity.
One can indeed identify two main sources for such differences.

The first one concerns the definition of the quantum heat current operator $\hat{\bm j}$, a well-known problem for the theoretical description of thermal transport  \cite{schelling2002comparison,marcolongo2016microscopic,carbogno2017ab}. Indeed, while in the case of the electrical current one can define the quantum current operator via the response to an external gauge field, in the case of the thermal current $\hat{\bm j}$ can only be defined via a continuity equation for the local energy density $\hat{h}$:
	\begin{equation}
		\label{continuity1}
		\dv{\hat{h}(\bm x,t)}{t}+\div  \hat{\bm j}(\bm x,t) = 0 \:.
	\end{equation}
However, the identification of the local energy density $\hat{h}$ is not unique, since the local partition of the total energy density of the crystal can be done in different ways. This non-uniqueness is reflected in the definition of the
heat current operator, and it has been used in Ref.\ \cite{ercole2016gauge} to formulate a so-called gauge-invariance principle for the thermal transport that is analogous to the well-known one for electrical current, \textit{i.e}.\ different definitions of the thermal current that lead to the same physical result for the transport coefficients. As we shall see, this issue is deeply connected with the approximation scheme used to compute the thermal conductivity. Among the recently published works, the Green-Kubo approaches of Refs.\  \cite{isaeva2019modeling,dangic2020origin}  relied on the original heat-flux definition proposed by Hardy  \cite{hardy1963energy},  where $\hat h({\bf x}, t)$ is essentially a coarse-grained version of the harmonic Hamiltonian expressed in terms of the ionic displacement operators and their conjugate momenta. 
On the other hand, some of us followed a different approach, where the heat flux is directly computed from a generalization of the BTE named Wigner transport equation (WTE)  \cite{simoncelli2019unified}, further detailed in Ref.\  \cite{long_paper}. This formulation exploits the Wigner transformation of the density matrix, which naturally suggests the introduction of bosonic operators which describe localized vibrational excitations. Even though in Refs.\  \cite{simoncelli2019unified,long_paper} the corresponding quantum operator for the heat current is not explicitly given, from the form of the matrix elements of phonon velocities presented in  \cite{simoncelli2019unified} one can already infer a different structure for it with respect to the one given by Hardy. 

A second apparent source of discrepancies between recent  \cite{pereverzev2018theoretical,isaeva2019modeling} and previous  \cite{maradudin1964lattice,semwal1972thermal} works based on the Green-Kubo formalism concerns the implementation of interaction effects for phonons. Indeed, while the original papers  \cite{maradudin1964lattice,semwal1972thermal} followed the standard route of implementing self-energy corrections in the frequency domain via a generalized phonon spectral function, in Ref.\  \cite{pereverzev2018theoretical,isaeva2019modeling} the calculations are done in the time-domain, {with approximated schemes needed to assign frequency-independent lifetimes to each phonon mode.} Even though the two approaches are expected to give the same final expressions in the limit where phonon broadening is sufficiently smaller than phonon energies, such a formal equivalence cannot be obtained, as we will see below, leaving open the question of a consistent derivation of thermal conductivity within the Green-Kubo approach.

The aim of the present manuscript is to derive a full quantum description of heat transport in the case of complex crystals {using the many-body Green's function approach},
{and} to explain analogies and differences between the various existing approaches, with the goal of pointing to a consistent theoretical framework for the computation of {thermal} conductivity in crystals. More specifically,  we derive the thermal conductivity in the so-called dressed-bubble approximation, where the effects of anharmonicity and disorder are encoded via renormalized phonon spectral densities including all self-energy corrections. 
The resulting expression of the thermal conductivity depends on the choice of the heat-flux operator.
Here we focus on two specific choices for the heat flux: the one originally proposed by Hardy  \cite{hardy1963energy}, and used later in several works  \cite{isaeva2019modeling,semwal1972thermal,dangic2020origin}, and the one obtained following the Wigner approach as proposed in Ref.\  \cite{long_paper}. In both cases, we provide a general expression able to describe thermal transport both in the low-damping regime, where phonon excitations are still well defined and interband transport is eventually relevant for densely spaced phonon bands, and in the overdamped regime, where the phonon quasiparticle picture breaks down. {To make a closer connection with the previous works, we show how some of the aforementioned results for thermal conductivity can be compared to our fully-quantum expression in the Lorentzian spectral function approximation (LSFA). The LSFA consists in approximating the imaginary part of the self-energy with its value computed at the (bare) phonon frequencies $\omega(\bm q)_s$ so that the spectral density reduces to a Lorentzian with full-width at half maximum $\Gamma(\bm q)_s$.} As we shall see, the two different heat fluxes yield conductivities differing exclusively in the contributions from interband processes. The heat flux derived by Hardy describes processes which are relevant for $\omega(\bm q)_s{-}\omega(\bm q)_{s'}{\lesssim} \Gamma(\bm q)_s{+}\Gamma(\bm q)_{s'}$, denoted as ``resonant'' in Ref.\   \cite{isaeva2019modeling}, and processes relevant for $\omega(\bm q)_s{+}\omega(\bm q)_{s'}{\lesssim} \Gamma(\bm q)_s{+}\Gamma(\bm q)_{s'}$, denoted as ``antiresonant''. We show that the former give a sizeable contribution to the thermal conductivity, while the latter are typically negligible, as expected since phonon frequencies are always positive. 
Within the Wigner formulation  \cite{simoncelli2019unified,long_paper} these antiresonant terms are absent. We show that using the Wigner heat flux in the LSFA yields exactly the result for the single-mode relaxation-time approximation (SMA) of the WTE reported in Ref.\  \cite{long_paper}.  

The recovery of the result for thermal conductivity presented in Ref.\  \cite{simoncelli2019unified} from the present quantum formula supports the theoretical consistency of the Wigner formalism for thermal transport. In the case of the Hardy current operator the present result corresponds to the ones given in Ref.\  \cite{maradudin1964lattice,semwal1972thermal}, but differs from more recent derivations  \cite{isaeva2019modeling,pereverzev2018theoretical}.  We show that such a discrepancy originates from the procedure used to implement the limit of frequency-independent phonon lifetimes within the real-time Green's function, and we show that this affects only the description of interband effects. 

Despite formal differences between the thermal-conductivity expressions derived from the Hardy or the Wigner heat-flux operators, the numerical results are quantitatively very similar in the regime where the LSFA holds. We investigate two test cases: perovskite \ch{CsPbBr3} and zirconate \ch{La2Zr2O7}, both materials with ultralow thermal conductivity. In these systems, we find that the different formulations for the heat flux lead to little or negligible numerical discrepancy, and both approaches give successful predictions in agreement with experimental data. Interestingly, we find that the same holds also for the results obtained via the real-time approximated scheme used in Refs.\  \cite{isaeva2019modeling,pereverzev2018theoretical}. So, the formal discrepancies in the final expressions of thermal conductivity become quantitatively irrelevant for the systems under investigation.  

We finally discuss the theoretical foundations for the consistency among the results obtained from the Hardy and Wigner heat fluxes, arguing that the gauge-invariance principle of transport coefficients  \cite{ercole2016gauge} might not be sufficient to explain such a similarity, and suggesting that a consistent theoretical prescription on how to choose the heat flux is still missing. In this regard, we provide some plausibility arguments which point toward the Wigner heat flux as the most appropriate choice. 


The paper is organized as follows: in Sec.\ref{sec:kubo formula} we show how to recast the Kubo formula for thermal conductivity in terms of the static limit of a finite-frequency response function, using the fluctuation-dissipation theorem. In Sec.\ref{sec:heat flux} we present the definitions of the local energy operator used in the Hardy  \cite{hardy1963energy} and in the Wigner  \cite{simoncelli2019unified} formalism, and we describe how to derive the corresponding heat-flux operators following Hardy's original procedure  \cite{hardy1963energy}. In Sec.\ref{sec:conductivity} we show how to compute thermal conductivity using the expressions that have been derived for the Hardy and Wigner heat-flux operators in the dressed-bubble approximation, which is analyzed as a perturbative diagrammatic expansion of finite-temperature Green's functions. Thereafter, the resulting expressions are discussed and compared, while addressing differences and analogies with the formulas already present in the literature  \cite{simoncelli2019unified,isaeva2019modeling,semwal1972thermal,maradudin1964lattice}.
Finally, in Sec.\ref{sec:discussion} we comment on the numerical results obtained in the LSFA, wondering if the agreement between the computed thermal conductivities obtained from the different definitions of the heat flux is a consequence of the gauge invariance of transport coefficients, and in Sec.\ref{sec:conclusion} we present some closing remarks.

\section{Kubo formula for thermal conductivity}
\label{sec:kubo formula}
The thermal conductivity $\kappa$ is defined as the tensor that relates the macroscopic heat flow in a system to the temperature gradient applied to it 
\begin{equation}
	\label{fourier}
	{ J}^{\a} {=} {-}\sum\limits_\b\kappa^{\a\b} \grad^{\b} T \:.
\end{equation}
In the linear-response regime the thermal conductivity at temperature $T$ is an equilibrium property, and as such can be computed in the absence of an external perturbation acting on the system using the Kubo formula for thermal conductivity, which reads  \cite{kubo2,mahan2013many,chester1963theory}
\begin{equation}
	\label{kubo lambda}
	\kappa^{\alpha\beta} {=} \frac{N_c\mathcal{V}}{ T} \lim\limits_{\epsilon{\to}0^{+}} \int\limits_0^\infty\!\!\dd{t} e^{{-}\epsilon t}\!\!\!\!\!\!\int\limits_0^{1/k_BT }\!\!\!\dd{\lambda} \langle\hat{J}^{\beta}(0)\hat{J}^{\alpha}(t{+}i\hbar\lambda)\rangle \:.
\end{equation}
This expression features the heat-flux operator $\hat{\bm J}$ as the volume integral of the microscopic current density $\hat{\bm{J}}{=}\frac{1}{N_c\mathcal{V}} \int_V\!\dd[3]{\!x} \hat{\bm{j}}(\bm{x}) $ (bold lettering indicates a cartesian vector) normalized by the volume of the sytem $V {=}N_c\mathcal{V}$, obtained multiplying the number of primitive cells $N_c$ by the primitive cell volume $\mathcal{V}$. In Eq.\ \eqref{kubo lambda} the time-evolution of the operators is intended in the Heisenberg representation while the angle brackets indicate the averaging in absence of thermal disturbances (\textit{i.e}.\ at equilibrium). Thus, for a generic operator $\hat O$ we have  $\hat O(t{+}i\hbar\lambda){=}e^{\frac{i}{\hbar}(t + i\hbar\lambda)\hat H }\hat Oe^{-\frac{i}{\hbar}(t + i\hbar\lambda)\hat H} $  where $\hat H$ is the unperturbed Hamiltonian of the system, while the expectation value $\langle\hat O\rangle$ is $ \Tr \small[ e^{-\frac{\hat H}{k_BT}}\hat O \small]/\mathcal{Z}$ with the canonical partition function $\mathcal{Z}{=}\Tr\small[e^{-\frac{\hat H}{k_BT}}\small]$. 
Note also that in the classical limit ($\hbar{\to}0$) Eq.\ \eqref{kubo lambda} reduces to the usual expression of the Kubo formula for thermal conductivity  as the time integral of the heat-flux auto-correlation function: $\kappa^{\alpha\beta}{=}\frac{N_c\mathcal{V}}{k_BT^2}\int\limits_0^\infty\!\!\dd{t}\langle{J}^{\alpha}(t){J}^{\beta}(0)\rangle$. The ordering of the cartesian components of the heat-flux operator does not matter, since the Onsager relations  \cite{onsager1931reciprocal} imply that $\k^{\a\b}{=}\k^{\b\a}$. In complete analogy with the standard approach to the calculation of electrical conductivity  \cite{mahan2013many}, Eq.\ \eqref{kubo lambda} can be expressed as the static limit of a finite-frequency response function. Let us consider the current-current correlation function, together with its Fourier transform (frequency integrals will always be intended from $-\infty$ to $+\infty$)
\begin{equation}
	\label{S = JJ}
	S^{\a\b}(t) = \langle \hat{J}^{\a}(t)\hat{J}^{\b}(0) \rangle = \int\!\! \dd{\w} e^{-i\w t}  S^{\a\b}(\w)\:;
\end{equation}
this is related to the retarded response function 
\begin{equation}
	\label{chi(t)}
	\chi^{\a\b}(t) = \frac{i}{\hbar}\theta(t)\langle[\hat{J}^{\a}(t),\hat{J}^{\b}(0)]\rangle
\end{equation}
by the fluctuation-dissipation theorem  \cite{kubo1966fluctuation} as follows
\begin{equation}
	\label{FD thm}
	S^{\a\b}(\w) = \frac{\hbar}{\pi}\bigr[n_T(\w) + 1 \bigl] \Im \chi^{\a\b}(\w + i0^+),
\end{equation}
where $n_T(\w) {=} ({e^{\frac{\hbar\w}{k_BT}}{-}1})^{-1}$ is the Bose-Einstein equilibrium distribution function at temperature $T$. Inserting \eqref{S = JJ} in \eqref{kubo lambda} and calculating the integrals in $\lambda$ and $t$ yields
\begin{equation}
	\label{k = s(w)}
	\k^{\a\b}{=} \frac{N_c\mathcal{V}}{ T}\lim\limits_{\epsilon\to 0^{+}}\int\dd{\omega}\frac{i}{\omega {+} i\epsilon} \frac{1{-}e^{-\frac{\hbar\omega}{k_BT}}}{\hbar \omega} S^{\a\b}(\omega) \:.
\end{equation}
At this point we can use Eq.\ \eqref{FD thm} in the latter equation and the well-known Sokhotski–Plemelj formula  \cite{abrikosov2012methods} $\lim\limits_{\epsilon\to 0^+}\frac{i}{\omega {\pm} i\epsilon} = i\mathcal{P}\frac{1}{\omega}{\pm}\pi\delta(\omega)$ to obtain
\begin{equation}
	\label{k = chi}
	\k^{\a\b}{=} \frac{N_c\mathcal{V}}{ T}\lim\limits_{\w \to 0} \frac{\Im \chi^{\a\b}(\w {+}i0^+)}{\w} \:.
\end{equation}
Note that when using the aforementioned result to solve the $\epsilon{\to} 0$ limit one actually gets an imaginary part of thermal conductivity related to a Cauchy principal value integral (denoted with $\mathcal{P}$). But if Onsager relations hold --- thus, if $\kappa$ is a symmetric tensor --- the expression \eqref{kubo lambda} is purely real. In fact, the $\epsilon{\to}0$ limits yields $\Im \k^{\a\b} {=} \frac{V}{k_BT}   \pv{\int \dd{\w} \frac{ \Im \chi^{\a\b}(\w+i0^+) }{\w^2}}$ which is zero since $ \Im\chi^{\a\b}(\w {+} i0^+){=}{-}\Im\chi^{\b\a}({-}\w {+} i0^+) $ as can be easily shown with a spectral decomposition of the response function.

The advantage of the representation \eqref{k = chi} over \eqref{kubo lambda} is that the imaginary part of the retarded response function can be computed as the analytic continuation of the time-ordered imaginary-time response function, as we will see in Sec. \ref{sec:conductivity}. The response function formalism is the most convenient choice to implement perturbative calculations of thermal conductivity when accounting for interactions. In contrast, the formulation \eqref{kubo lambda} is in practice feasible only in the so-called bubble approximation, as we will discuss in more detail in what follows and in the Appendix \ref{app:conductivity}.

\section{Heat-flux operator for a crystal lattice}
\label{sec:heat flux}
In this Section we  describe how to derive the heat-flux operators following Hardy's original procedure  \cite{hardy1963energy} starting from the definitions of the on-site energy operator used in the Hardy  \cite{hardy1963energy} and in the Wigner  \cite{simoncelli2019unified} formalism. In fact, to calculate the thermal conductivity from \eqref{k = chi}, a definition of the heat-flux operator for a lattice must be introduced. The standard picture of heat transport in crystals relies on the Peierls-Boltzmann transport equation in its linearized form (LBTE)  \cite{peierls1929kinetischen}, in which phonon interband effects are neglected. This model pictures the crystal lattice as a semiclassical gas of propagating phonon wavepackets that scatter like particles, and {in} this formalism, the heat-flux operator $\hat{\bm{J}}_P$ accounts for particle-like propagation via independent phonon bands 
\begin{equation}
	\label{J pbte}
	\hat{\bm J}_{P}{=}\frac{1}{N_c\mathcal{V}}\sum\limits_{\bm q,s} \h\omega(\bm q)_{s} \bm v(\bm q)_s \hat{n}(\bm q)_{s},
\end{equation}
where $\omega(\bm q)_s$ and $\bm v(\bm q)_s{=}\grad_{\bm q} \w(\bm q)_s$ are respectively the frequency and group velocity of phonon with quasi-momentum $\bm q$ and in band $s$, and the sum in $\bm q$ spans over the values of crystal momentum fixed by periodic boundary conditions contained in the first Brillouin zone. The heat flux is diagonal in the phonon band index, depending only on the phonon number operator  $\hat{n}(\bm q)_{s}{=}\hat a^\dagger\!(\bm q)_s \hat a(\bm q)_{s}$, where $\hat{a}^\dagger\!(\bm q)_s $ ($ \hat{a}(\bm q)_s $) denote the creation (annihilation) operator of a phonon with wavevector $\bm q$ belonging to the band $s$  \cite{ziman2001electrons,wallace1972thermodynamics}. Within the SMA, each phonon mode contributes individually to the lattice thermal conductivity, which reads  \cite{ziman2001electrons}
\begin{equation}
	\label{k pbte}
	\kappa_P^{\a\b} {=}\frac{1}{N_c\mathcal{V}}\sum\limits_{\bm q, s} C(\bm q)_{s} v^\a(\bm q)_s v^\b(\bm q)_s \tau(\bm q)_s
\end{equation}
where $C(\bm q)_{s} {=}\frac{\hbar^2 \w^2(\bm q)_s}{k_B T^2} n_T(\w(\bm q )_s) [n_T(\w(\bm q )_s) {+} 1]$ is the contribution of the mode $(\bm q, s)$ to the heat capacity of the lattice and $\tau(\bm q)_s{=}\tfrac{1}{\Gamma(\bm q)_s}$ is the phonon lifetime.  Eq.\ \eqref{k pbte}  basically describes the thermal conductivity of a gas  \cite{allen1993thermal} of particles labeled by $(\bm q, s)$, each characterized by specific heat  $C(\bm q)_{s}$, velocity $\bm v(\bm q)_s $ and lifetime $\tau(\bm q)_s $. As we shall see below, Eq.\ \eqref{k pbte}  consist in the intraband contribution to the thermal conductivity --- \textit{i.e}.\ the SMA limit for $\kappa_P$ in Ref.\  \cite{long_paper}--- in the LSFA of the result derived from the fully quantum formula \eqref{k = chi}.

Here we want to describe thermal transport beyond Peierls-Boltzmann, \textit{i.e}.\ accounting not only for intraband propagation but also for interband effects. Thus, a generalized definition of the heat flux that is non-diagonal in the phonon band indices is needed. In fact, while in normal crystals with well-separated and defined phonon bands  diagonal heat-flux elements are dominant (since $\langle \hat a^\dagger(\bm q)_s \hat a(\bm q)_{s}\rangle{\gg} \langle \hat a^\dagger(\bm q)_s \hat a(\bm q)_{s\p}\rangle $ for $s{\neq}s\p$), when the phonon bands become close in energy phonons can not only propagate particle-like but also tunnel from one band to another other in a wave-like fashion. This happens in glasses and disordered solids, where the vibrational spectrum is quite dense featuring many quasi-degenerate modes  \cite{allen1993thermal,feldman1993thermal}, and in complex crystals, where strong anharmonicity broadens the phonon bands and provides overlap between different states. In these cases, interband transitions are of paramount importance, giving significant or even dominant contributions to heat transport  \cite{wang2018cation,lee2017ultralow,pisoni2014ultra}. 
Interband effects are absent from the equations \eqref{J pbte}-\eqref{k pbte} derived from the LBTE, and in the following, we illustrate how these emerge from a wave-like interference between different vibrational modes. This physical picture is further detailed in Appendix \ref{app:greens functions}. 

The Hamiltonian that describes ionic motion for a crystal within the harmonic approximation reads
\begin{equation}
	\label{Hamiltonian}
	\hat{\mathcal H}{=} \sum\limits_{\textbf{R}b\a} \frac{\hat{p}^2(\bm R)_{b\a}}{2M_b} {+} \frac{1}{2} \sum\limits_{\substack{\textbf{R}b\a\\\textbf{R}^\prime \!b^\prime\!\a^\prime }} \bm \Phi_{\bm Rb\a, \bm R^\prime b^\prime \a^\prime} \hat{u}(\bm R)_{b\a}\hat{u}(\bm R\p)_{b\p\!\a\p},
\end{equation}
where the sum on $\bm R$ indicates the sum over the $N_c$ primitive cells located at lattice sites  $\bm R$, the index $b{=} 1,\dots, N_{at}$ runs over all the atomic species inside the primitive cell of volume $\mathcal{V}$ and $\alpha$ denotes a cartesian direction. The equilibrium positions of the atoms inside the primitive cell are identified by the vectors $\bm{\tau}_b$, \textit{i.e}.\ the equilibirum position of atom $b$ inside the primitive cell $\bm R$ is $  \bm R{+}\bm{\tau}_b$. The ionic displacement from the equilibrium position $\hat{u}$ and ionic momentum $\hat{p}$ are canonically conjugated variables satisfying the usual commutation rules
\begin{equation}
	\label{[u,p]}
	[\hat{u}(\bm R)_{b\a},\hat{p}(\bm R^\prime)_{b^\prime\a^\prime}] = i \h \delta_{\bm R,\bm R^\prime} \delta_{b,b^\prime} \delta_{\a,\a^\prime}\:.
\end{equation}
The interatomic force constant matrix $\bm \Phi$ is defined as the the second derivative of the Born-Oppenheimer potential with respect to ionic displacements, which is symmetric and translational invariant  \cite{peierls1955quantum}
\begin{equation}
	\label{prop phi}
	\bm \Phi_{\bm Rb\a, \bm R^\prime b^\prime \a^\prime} = \bm \Phi_{ \bm R^\prime b^\prime \a^\prime, \bm Rb\a} =\bm \Phi_{(\bm R- \bm R\p)b\a, \bm 0 b^\prime \a^\prime}  \:.
\end{equation}
In our framework, we will consider the harmonic Hamiltonian \eqref{Hamiltonian} as the unperturbed Hamiltonian of the system, whereas the higher-order anharmonic terms will be considered as perturbations.

In order to derive the heat-flux operator, we follow the procedure originally derived by Hardy \cite{hardy1963energy}. The starting point is the identification of a local energy field operator $\hat h(\bm R)_{b\a}$ that satisfies 
\begin{equation} \label{extensivity}
	\hat{\mathcal H} = \sum_{\bm Rb\a} \hat{h}(\bm R)_{b\a}\:.
\end{equation}
From there, the expression of the heat-flux operator is derived using the continuity equation 
\begin{equation}
	\label{continuity1}
	\dv{\hat{h}(\bm x,t)}{t}+\div  \hat{\bm j}(\bm x,t) = 0 \:.
\end{equation}
and relying on the assumption of being in the close-to-equilibrium regime, where temperature variations are appreciable over a length scale much larger than the interatomic spacing. The identification of $\hat h(\bm R)_{b\a}$ from the structure of the Hamiltonian is non-trivial. In fact, there are an infinite number of ways to partition energy while satisfying the requirement \eqref{extensivity}. As a consequence of this arbitrariness, the resulting heat-flux operator is not uniquely defined. The definition of the generalized heat flux which is most commonly adopted is the one derived originally by Hardy  \cite{hardy1963energy}, which has been successfully implemented to calculate interband thermal transport phenomena in recent studies  \cite{pereverzev2018theoretical,isaeva2019modeling,dangic2020origin}. On the other hand, in Ref.\  \cite{simoncelli2019unified} a different definition of heat-flux operator arises within the theoretical framework of the WTE. In the present section, the two different definitions of the heat-flux operator are compared and analyzed.

Nonetheless, as shown in Ref.\  \cite{ercole2016gauge}, whenever two choices of the microscopic energy density differ by the divergence of a bounded vector field, they realize two equivalent ``gauges'' of the energy field and lead to the same result for the thermal conductivity (further discussion in Appendix \ref{app:gauge}). This gauge-invariance principle cannot be invoked, though, for local energies which do not differ by a divergence. We argue that this is the case for the two heat fluxes we discuss here, that do not seem to be connected by a gauge transformation in the sense of Ref.\  \cite{ercole2016gauge}, even though the numerical results for thermal conductivity are quantitatively very close. We stress that both definitions of the local energy lead to harmonic heat fluxes, but with a different structure for the interband contribution, as alluded to above. 

\subsection{Hardy heat flux} \label{subsec:heat flux hardy}

In this Section, we present the procedure to derive the heat-flux operator originally proposed by Hardy in Ref.\  \cite{hardy1963energy}. The starting point is the definition of the local energy field; the one used by Hardy is the most natural choice given the structure of the Hamiltonian \eqref{Hamiltonian}
\begin{equation}
	\label{hardy h(R)}
	\hat{h}(\bm R)_{b\a} {=} 
	\frac{\hat{p}^2(\bm R)_{b\a}}{2M_{b}} {+}\frac{1}{2}\!\sum_{\bm{R}\p\! b\p\! \a\p} \!\bm{\Phi}_{\bm{R}b\a, \bm{R}\p b\p \a\p} 
	\hat{u}(\bm R)_{b\a}\hat{u}(\bm R\p)_{b\p\!\a\p} .
\end{equation}
In order to derive the heat-flux operator from the continuity equation \eqref{continuity1} one needs to introduce an energy density operator $\hat h(\bm x)$, which is a continuous function of the space coordinate  $\bm x$, such that its integral in space gives the harmonic Hamiltonian \eqref{Hamiltonian}. This is done by convoluting the energy field operator \eqref{hardy h(R)} with a smooth normalized distribution  $\Delta_\ell(\bm x{-}\bm R{-} \bm{ \bm{\tau}}_b)$ centered on the ionic equilibrium position $\bm R +\bm{\tau}_b$ and spatially spread over a mesoscopic length $\ell$ (\textit{i.e}.\ $\ell{\gg}a$ where $a$ is the microscopic bond length), such as $\Delta_\ell(\bm x{-}\bm R{-} \bm{ \bm{\tau}}_b){=} \frac{1}{\ell^3\sqrt{\pi^3}} \exp[-\frac{|\bm x -\bm R - \bm{\bm{\tau}}_b|^2}{\ell^2}] $. From \eqref{hardy h(R)}, the energy density operator is defined as 
\begin{equation}
	\label{hardy h(x)}
	\hat{h}(\bm x) = \sum\limits_{\bm R b \a} \hat{h}(\bm R)_{b\a} \Delta_\ell (\bm x {-} \bm R {-} \bm{\bm{\tau}}_b ), 
\end{equation}
and it is easy to show that the normalization of the $\Delta_\ell$ function ensures $\hat{\mathcal H} {=} \int_V\!\dd[3]{x}  \hat{h}(\bm x)$. Such a coarse-graining procedure on a scale $\ell$ implicitly assumes that the temperature gradient varies on scales larger than $\ell$, so that the energy density at a given point $\bm x$ is (mostly) built from the energy of the ions contained in a domain of linear dimension $\ell$  \cite{hardy1963energy,allen1993thermal}. .The heat-flux operator can be now derived from \eqref{hardy h(x)} using the continuity equation \eqref{continuity1} and the {Heisenberg time-evolution equation} for the energy density in the harmonic approximation:
\begin{equation}
	\label{continuity}
	-\div  \hat{\bm j}(\bm x) = \dv{\hat{h}(\bm x)}{t} = -\frac{i}{\h} [\hat{h}(\bm x),\hat{\mathcal H}]  \:.
\end{equation}
The total heat-flux operator is then obtained by inserting \eqref{hardy h(x)} in \eqref{continuity}, performing a Taylor expansion of the function $\Delta_\ell(\bm x{-} \bm R {-} \bm{\bm{\tau}}_b)$, and finally integrating the heat current density over the crystal volume and considering only the contribution from the lowest-order term of the aforementioned Taylor expansion. This procedure, outlined in  Appendix \ref{app:heat flux}, yields 
\begin{equation}
	\begin{split}
		\label{J = up}
		\hat{\bm J}& = -\frac{i}{N_c\mathcal{V}\hbar} \sum\limits_{\bm R b \a}( \bm R + \bm{\bm{\tau}}_b) [\hat{h}(\bm R)_{b\a}, \hat{\mathcal H}] \\
		{=}\frac{1}{2N_c\mathcal{V}}\!\!\sum\limits_{\substack{\bm Rb\a \\\bm R\p b\p\!\a\p} }&\!\!(\bm R{+}\bm{\bm{\tau}}_b{-} \bm R\p{-}\bm{\bm{\tau}}_{b\p} )  \bm \Phi_{\bm Rb\a, \bm R\p b\p \a\p} \hat{u}(\bm R)_{b \a}  \frac{\hat{p}(\bm R\p)_{b\p\!\a\p}}{M_{b\p}} .
		\raisetag{22mm}
	\end{split}
\end{equation}
 Notice that, since we are interested in the heat flux $\hat{\bm{J}}$ which is the volume average of the current density $\hat{\bm j}(\bm x)$, the microscopic details of the coarse-graining procedure are irrelevant, as testified by the fact that the expression of the heat flux reported above does not feature the characteristic length $\ell$ introduced to define the local energy density operator $\hat h (\bm x)$ in Eq.\ \eqref{hardy h(x)}. 
 
The total heat flux~\eqref{J = up} has been derived in real space relying on the assumption of being in the close-to-equilibrium regime, which is required to employ the momentum $\hat{p}(\bm R)_{b\a}$ and displacement $\hat{u}(\bm R)_{b \a}$ operators of atoms at well-defined positions in real space $\bm{R}{+}\bm{\tau}_b$. 
However, the evaluation of its expectation value does not necessarily require to employ a real-space representation, and for later convenience we recast it in reciprocal space, where translation-invariant quantities assume a block-diagonal form. To see this, let us recast Eq.\ \eqref{J = up} employing the standard phonon annihilation (creation) operators $\hat{a}(\bm q)_{s}$ ($\hat{a}^\dagger(\bm q)_{s}$),  which are labeled by a wavevector $\bm q$ belonging to the first Brillouin zone and a mode index $s$  \cite{ziman2001electrons,wallace1972thermodynamics}, and are related to the displacement and momentum operators via
\begin{equation}
	\begin{split}
		\raisetag{18mm}
		\hat{a}(\bm{q})_{s}
		&{=}\frac{1}{\sqrt{N_c}} \sum_{\bm{R},b,\alpha}\!\mathcal{E}^{\star}\!(\bm{q})_{s,b\alpha}\Bigg[
		\frac{{\hat{p}}(\bm{R})_{b\alpha}}{\!\sqrt{2 \hbar\omega(\bm{q})_sM_b}} \\
		&\hspace{2cm}-i \frac{\sqrt{\omega(\bm{q})_sM_b}}{2 \hbar}
		{\hat{u}}(\bm{R})_{b\alpha}
		\Bigg]e^{-i\bm{q}\cdot(\bm{R}+\bm{\tau}_b)}.
	\end{split}
	\label{eq:ph_annihilation}
\end{equation}
In Eq.\ (\ref{eq:ph_annihilation}), $\mathcal{E}(\bm q)_{s,b\a}$ are the phonon polarization vectors and $\w(\bm q)_s$ are the phonon frequencies.
These quantities are obtained from the dynamical matrix
\begin{equation}
	\label{dynmat}
	\tenscomp{D}(\bm q)_{b\a,b\p\!\a\p} = \sum\limits_{\bm R} 
	\frac{\bm \Phi_{\bm R b\a,\bm R \p b\p\!\a\p}}{\sqrt{M_bM_{b\p}}}
	e^{-i \bm q \vdot (\bm R + \bm{\tau}_b- \bm{\tau}_{b\p})}
\end{equation}
by solving the eigenvalue equation $\w^2(\bm q)_s\mathcal{E}(\bm q)_{s,b\a}{=}\sum_{b\p\a\p}\tenscomp{D}(\bm q)_{b\a,b^\prime\a\p} \mathcal{E}(\bm q)_{s,b^\prime\a\p}$. Combining Eq.\ (\ref{eq:ph_annihilation}) and Eq.\ (\ref{[u,p]}), it is straightforward to show that the phonon creation and annihilation operators satisfy the usual bosonic commutation rules 
\begin{equation}
	\label{[a,a]}
	[\hat{a}(\bm q)_s,\hat{a}^\dagger(\bm q\p)_{s\p}  ] =  \delta_{s,s\p} \delta_{\bm q , \bm q\p} \:.
\end{equation}
The usefulness of these phonon operators is to simplify the expression for the harmonic Hamiltonian \eqref{Hamiltonian} to a diagonal form
\begin{equation}
	\label{H = aqs aqs}
	\hat{\mathcal H} =\!\!\sum\limits_{\bm q,s}\hbar \omega(\bm q)_s \bigr[ \hat{a}^\dagger(\bm q)_s\hat{a}(\bm q)_{s} + \tfrac{1}{2} \bigl] \:,
\end{equation}
as well as that of the heat flux~\eqref{J = up} to a block-diagonal form
\begin{equation}
	\begin{split}
		\bm{\hat{J}} {=} \frac{1}{N_c\mathcal{V}}\sum_{\bm q,ss\p} \h \omega(\bm q)_{s} \bm v(\bm q)_{ss^\prime} \tfrac{1}{2}\left[\hat a^\dagger(\bm q)_s{+}\hat a(-\bm q)_s \right] \\  \times \left[\hat a(\bm q)_{s^\prime}{-}\hat a^\dagger(-\bm q)_{s^\prime} \right],   \label{J hardy}
		\raisetag{-10mm}
	\end{split}
\end{equation}
where the generalized phonon velocity $\bm v(\bm q)_{ss^\prime}$ is a $3N_{at}{\times}3N_{at}$ matrix for every Cartesian component, defined as
\begin{equation}
	\label{velocity hardy}
	\bm v(\bm q)_{ss\p} {=}\tfrac{1}{2\sqrt{\w(\bm q)_{s}\w(\bm q)_{s\p} }} \mathcal{E}^*(\bm q)_{s,b\a} \grad_{\bm q}\tenscomp{D}(\bm q)_{b\a,b\p\!\a\p}\mathcal{E}(\bm q)_{s\p,b\p\!\a\p} \:.
\end{equation}
The diagonal $s{=}s\p$ elements of this matrix are the usual phonon group velocities $\bm v(\bm q)_{ss}{=}\grad_{\bm q} \w(\bm q)_s$. Other properties of the velocity matrix \eqref{velocity hardy} are
\begin{equation}
	\label{vel properties}
	\begin{split}
		\bm v(\bm q)_{ss\p}&{=}\bm v^*(\bm q)_{s\p\! s}{=}{-}\bm v(-\bm q)_{s\p\! s} 
	\end{split}
\end{equation}
These relations follow directly from the properties of the dynamical matrix \eqref{dynmat}, which are consequence of the symmetries \eqref{prop phi} of the interatomic force constants matrix. Note that the phase for the Fourier transform in Eqs.\eqref{eq:ph_annihilation},\eqref{dynmat} features the atomic equilibrium position $e^{-i \bm q \vdot (\bm R + \bm{ \bm{\tau}}_b )}$ (``Wallace convention''  \cite{wallace1972thermodynamics})  instead of just the primitive cell position $e^{-i \bm q \vdot \bm R } $ (``Ziman convention''  \cite{ziman2001electrons}).
Employing the Ziman or Wallace phase convention does not affect the eigenvalues of the dynamical matrix, but it can change the off-diagonal elements of the velocity matrix. The Wallace phase convention for the Fourier transform possesses the advantage of being independent of the choice of the crystal's unit cell and of yielding a thermal conductivity expression that is size consistent; further discussions and details are presented in Ref.\  \cite{long_paper}. 

Using the properties of the velocity matrix and phonon frequencies and considering the commutation rules \eqref{[a,a]}, we can rearrange the Hardy heat-flux operator as the sum of a ``resonant'' heat-flux operator $\hat{\bm J}_R$ and a ``antiresonant'' heat-flux operator  $\hat{\bm J}_A$, according to the terminology introduced in Ref.\  \cite{isaeva2019modeling}. They are defined as
\begin{subequations}
	\begin{eqnarray}
		&& \bm{\hat{J}} \: = \bm{\hat{J}}_R + \bm{\hat{J}}_A,   \\
		\label{J res}
		&&\bm{\hat{J}}_R {=}\frac{\hbar}{N_c\mathcal{V}}\sum_{\bm q,ss\p} \frac{\omega(\bm q)_{s} \!+\! \omega(\bm q)_{s\p}}{2}  \bm v(\bm q)_{ss^\prime} \hat a^\dagger\!(\bm q)_s \hat a(\bm q)_{s^\prime},\quad \\
		&&\begin{split}
			\bm{\hat{J}}_A {=}\frac{\hbar}{N_c\mathcal{V}}&\sum_{\bm q,ss\p}\frac{\omega(\bm q)_{s}\! -\! \omega(\bm q)_{s\p}}{4} \bm v(\bm q)_{ss^\prime} \\ 
			&\hspace{.6cm} \times\left[\hat a(-\bm q)_s \hat a(\bm q)_{s^\prime}\!-\!\hat a^\dagger(\bm q)_s\hat a^\dagger(-\bm q)_{s^\prime} \right]\! \,.
		\end{split}    \label{J ares}
	\end{eqnarray}
\end{subequations}
\begin{figure}[h]
	\centering
	\includegraphics[width=0.5\textwidth,keepaspectratio]{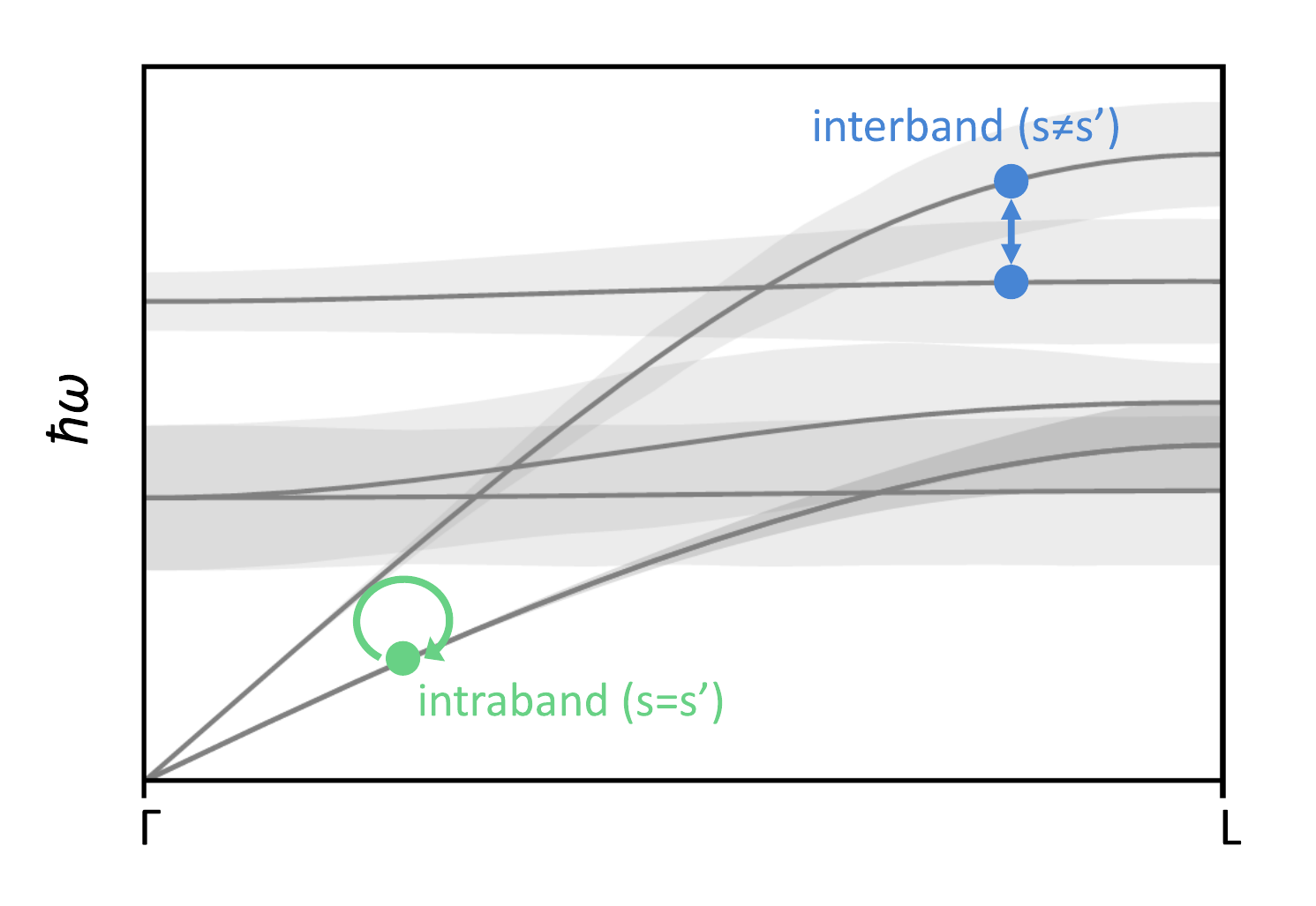}
	\caption{ Pictorial representation of the intraband and interband transitions described by Hardy's resonant \eqref{J res} and Wigner \eqref{J WBTE} heat-flux operators. Solid lines represent phonon bands of a representative crystal, shaded areas represents finite phonon linewidths. The diagonal elements ($s{=}s\p$) of the heat flux describe the particle-like intraband propagation of phonon wavepackets. The off-diagonal ($s{\neq}s\p$) elements describe the interband tunneling of phonons between two different bands: while as $\Gamma(\bm q)_s{\to}0$ these processes are only possible when an external perturbation provides the finite energy difference $\hbar\w_\Delta {=}\hbar \w(\bm q)_s {-} \hbar\w(\bm q)_{s\p}$; as $\Gamma(\bm q)_s$ increases, the tunneling between the two states is possible  even at $\hbar\w_\Delta {=}0$. 
	}
	\label{fig:bands-scheme}
\end{figure}
The resonant combination of operators $a^\dagger(\bm q)_{s}a(\bm q)_{s\p}$  gives us a physical picture of a vertical transition (tunneling at equal crystal momentum $\bm q$) in which a mode $s\p$ is destroyed and a mode $s $ is created (see Fig.\ \ref{fig:bands-scheme} for a pictorial interpretation). In absence of degeneracies, the diagonal elements $s'{=}s$ represents a transition within the same band, while for $s'{\neq} s$ the transition involves different bands.
The expectation value of the operators that realize the intraband transition $\langle a^\dagger\!(\bm q)_s a(\bm q)_s\rangle{=}n(\bm q)_s$ can be interpreted as the population of the phonon mode $(\bm q, s)$, with definite energy, momentum and band, which corresponds to a particle-like excitation with precise nature. Instead, the interband terms $\langle a^\dagger\!(\bm q)_s a(\bm q)_{s\p}\!\rangle$ describe the coherence between pairs of eigenstates, \textit{i.e}.\ coupling between two different vibrational modes $s,s\p$ with the same wavevector $\bm q$, and cannot be directly interpreted as a vibrational excitation of definite nature.
When two bands are degenerate, \textit{i.e}.\ $\omega(\bm q)_{s}{=}\omega(\bm q)_{s'}$ for $s'{\neq}s$, one can exploit the freedom of diagonalizing at least one Cartesian component of the velocity operator in the degenerate subspace  \cite{fugallo2013ab,long_paper}, obtaining a zero interband contribution from perfectly degenerate vibrational modes and an intraband conductivity along such Cartesian direction exactly equivalent to the Peierls-Boltzmann conductivity discussed in Ref.\ \cite{fugallo2013ab}. For these reasons we employ the convention of considering interband contributions to the conductivity to emerge exclusively from off-diagonal and non-degenerate velocity-operator elements. 
Since the temperature gradient is constant in time, a finite-frequency transition would not be allowed by the external perturbation, as evidenced by the fact that the thermal conductivity is defined as the $\omega{\rightarrow}0$ limit of the current-current correlation function (see Eq.\ \eqref{k = chi}). As a consequence, a perfect crystal with no phonon damping $\Gamma(\bm q)_s{=}0, \: \forall (\bm q ,s)$ has infinite thermal conductivity, since no relaxation process can hinder the intraband thermal conduction  (cf. Eq.\ \eqref{k pbte}). As soon as phonon states get broadened by anharmonicity or disorder ($\Gamma(\bm q)_s{\neq}0$), intraband transport becomes limited to a finite value and the additional interband transport mechanisms become possible \cite{nota_FPU}. As mentioned in the introduction, this will be relevant whenever broadening of the phonon energy levels overcomes their energy difference, \textit{i.e}.\ $ \Gamma(\bm q)_s{+}\Gamma(\bm q)_{s'}{\gtrsim}\omega(\bm q)_s{-}\omega(\bm q)_{s'}$, as pictorially shown in Fig.\ \ref{fig:bands-scheme}. The antiresonant combination $a(\bm q)_{s}a(\bm q)_{s\p}$ represents some sort of  ``pair tunneling'' process, typical in the context of superfluidity  \cite{deo1966calculation}, where a condensate reservoir exists. However, in the case of phonons, there is no condensation so these processes are expected to be irrelevant, as indeed we will confirm via both analytical and numerical estimates.

We finally note that in the original work by Hardy  \cite{hardy1963energy} the heat-flux operator had also anharmonic terms. Such a difference is due to the choice done in Ref.\  \cite{hardy1963energy} of centering the $\Delta_\ell$  function of Eq.\ \eqref{hardy h(x)} on the instantaneous position of the ions $\bm R{+}\bm{\bm{\tau}}_b {+} \hat{\bm {u}}(\bm R )_b$ rather than just on their equilibrium position, as we did. This would introduce anharmonic $\mathcal{O}( p^3) $ (convective) and $\mathcal{O}( u^3) $ (conductive) terms in the energy density, even if calculated from the harmonic on-site energy \eqref{hardy h(R)}. As shown in Appendix \ref{app:gauge}, the resulting heat-flux operator differs by a total time derivative from the one given in Eqs.\ \eqref{J res}-\eqref{J ares} (if mass diffusion is negligible  \cite{isaeva2019modeling}). Thus, owing to the gauge invariance principle for heat transport  \cite{ercole2016gauge},  this term does not affect the total thermal conductivity of the system when the exact correlation function \eqref{S = JJ} is computed. However, when the correlation function \eqref{S = JJ} is computed within an approximation, such as the dressed-bubble approximation employed in this work, this additional term cannot be neglected \textit{a priori}. 

Nonetheless, as already argued in Ref.\  \cite{hardy1963energy}, at the level of the present approximation the anharmonic terms in the heat current lead to higher-order corrections; we then expect a negligible contribution to the thermal conductivity and will retain in what follows only the harmonic terms \eqref{J res}-\eqref{J ares}. 

\subsection{Wigner heat flux} \label{subsec:heat flux wig}
In this Section, we show how the procedure outlined in Sec.\ref{subsec:heat flux hardy} for the derivation of the heat flux can be employed to obtain a new expression of the quantum-mechanical heat-flux operator which is in close connection with the Wigner formulation of thermal transport, developed in Refs.\  \cite{simoncelli2019unified,long_paper}. 
The starting point is once again the identification of the local energy operator. The local energy operator \eqref{hardy h(R)} follows quite naturally from the expression of the harmonic Hamiltonian \eqref{Hamiltonian} in the representation of atomic momentum and displacement operators $\hat p $ and $\hat u$. Now let us show how a different representation of the harmonic Hamiltonian provides a different yet {equally} valid choice of the local energy, from which a different heat-flux operator follows. In Ref.\  \cite{long_paper} the bosonic operators in real space were introduced as a set of creation (annihilation) operators that describe the excitation (de-excitation) of space-dependent atomic vibrations, and they were shown to be suitable to describe an out-of-equilibrium solid with space-dependent vibrational energy. These operators are related to the atomic momentum and displacement operators via
\begin{equation}
	\label{cartesian operators a(R)}
	\begin{split}
		\hat{\tenscomp{a}}(\bm R )_{b\a} {=} \frac{1}{\sqrt{2\h}}\sum_{\bm R\p\! b\p\! \a\p}\big( \sqrt[4]{\tenscomp{G}^{-1}\!\!}_{\bm R b\a,\bm R\p b\p\! \a\p} \frac{\hat{p}(\bm R\p)_{b\p \a\p}}{\sqrt{M_{b\p}}} \\
		-i \sqrt{M_b}\sqrt[4]{\tenscomp{G}}_{\bm R b\a,\bm R\p b\p\! \a\p} \hat{u}(\bm R\p)_{b\p\!\a\p} \big),
	\end{split}
\end{equation}
where the n-th root of the matrix $\tenscomp{G}_{\bm R b\a, \bm R \p b\p \a \p}{=}\tfrac{\bm \Phi_{\bm R b\a,\bm R \p b\p\!\a\p}}{\sqrt{M_bM_{b\p}}}$ is obtained as the inverse Fourier transform of the n-th root of the dynamical matrix \eqref{dynmat}, 
$\sqrt[n]{\tenscomp{G}}_{\bm R b\a,\bm R \p b\p\!\a\p}{=}\frac{1}{N_c}\sum\limits_{\bm q} \sqrt[n]{\tenscomp{D}(\bm q)}_{b\a,b\p\!\a\p}e^{+i \bm q \vdot (\bm R + \bm{\tau}_b- \bm{R'}-\bm{\tau}_{b\p})}$, and $\sqrt[n]{\tenscomp{\tenscomp{D}}(\bm q)}_{b\a,b\p\!\a\p}{=}\sum_s\w^{\frac{2}{n}}(\bm q )_s \mathcal{E}(\bm q)_{s,b\a} \mathcal{E}^*(\bm q)_{s,b\p\a} $. One can easily show from Eq.\ \eqref{cartesian operators a(R)} that these operators satisfy the usual bosonic commutation rules
\begin{equation}\label{[a(R),a(R)]}
	[\hat{\tenscomp{a}}(\bm R )_{b\a},\hat{\tenscomp{a}}^\dagger(\bm R^\prime )_{b^\prime\a^\prime} ] = \delta_{\bm R,\bm R^\prime} \delta_{b,b^\prime} \delta_{\a,\a^\prime},
\end{equation}
which follow trivially from the commutation rules \eqref{[a,a]} and the properties of the phonon polarization vectors. 
Ref.~ \cite{long_paper} shows how the operator $\hat{\tenscomp{a}}^\dagger(\bm R )_{b\a}$ creates atomic vibrations along direction $\alpha$ and centered around the position $\bm R+\bm{\tau}_b$, thus it can be used (together with its adjoint) to describe space-dependent vibrations. 

In terms of the bosonic operators in real space, the harmonic Hamiltonian reads
\begin{equation}
	\label{H = a(R) a(R)}
	\hat{\mathcal{H}}{=}\h\!\!\sum_{\substack{\bm R b\a \\ \bm R\p\! b\p\! \a\p}}\!\!\!\! \sqrt{\tenscomp{G}}_{\bm R b\a,\bm R\p b\p\!\a\p}\!\bigl[ \hat{\tenscomp{a}}^\dagger(\bm R)_{b\a}\hat{\tenscomp{a}}(\bm R\p)_{b\p\!\a\p} {+} \frac{1}{2}\delta_{\bm R,\bm R\p}\delta_{b,b\p}\delta_{\a,\a\p}\bigr].
\end{equation}
From Eq.\ (\ref{H = a(R) a(R)}), the local energy operator is naturally defined as
\begin{equation} \label{e(R) WBTE}
	\hat{\tenscomp{h}}(\bm R)_{b\a} {=} \frac{\h}{2}\!\!\sum_{\bm R\p\! b\p\!\a\p}\!\!\!\!\sqrt{\tenscomp{G}}_{\bm R b\a,\bm R\p b\p\!\a\p}\hat{\tenscomp{a}}^\dagger(\bm R)_{b\a}\hat{\tenscomp{a}}(\bm R\p)_{b\p\!\a\p} {+}\text{h.c.} \:,
\end{equation}
where the hermitian conjugate (h.c.)\ is added to enforce hermiticity of the local energy and heat-flux operators, whereas the zero-point energy term can be neglected since as a constant it cannot yield any heat flux. Given the definition of the local energy operator \eqref{e(R) WBTE}, we can apply the same procedure outlined in the previous Section to derive the heat-flux operator. As detailed in Appendix \ref{app:heat flux}, its expression in terms of the standard phonon operators is  
\begin{equation}
	\label{J WBTE}
	{\hat{\tens{J}}} {=} \frac{\hbar}{N_c\mathcal{V}}\sum_{\bm q,ss\p}\frac{\omega(\bm q)_{s}\! +\! \omega(\bm q)_{s\p}}{2}   \tens{v}(\bm q)_{ss^\prime} \hat a^\dagger(\bm q)_s \hat a(\bm q)_{s^\prime}
\end{equation}
where the velocity matrix is now defined as 
\begin{equation}
	\label{velocity WBTE}
	\tens{v}(\bm q)_{ss^\prime} =  \mathcal{E}^*(\bm q)_{s,b\a} \grad_{\bm q}\! \sqrt{\tenscomp{\tenscomp{D}}(\bm q)}_{b\a,b\p\!\a\p}\mathcal{E}(\bm q)_{s\p,b\p\!\a\p} \:.
\end{equation}
We will refer to this operator as the Wigner heat-flux operator, and accordingly to the local energy \eqref{e(R) WBTE} as the Wigner local energy operator. This nomenclature is chosen to underline the analogies with the heat flux discussed in Wigner's phase-space formulation for thermal transport presented in Ref.\  \cite{long_paper}. In fact, in the Wigner approach the heat flux is derived considering the time evolution of a local energy field $E(\bm R, t){=} \frac{\hbar}{\mathcal{V}N_c}\sum_{qs} \omega(\bm q)_s\tenscomp{n}(\bm R,\bm q,t)_{ss}$ where the matrix $\tenscomp{n}(\bm R,\bm q,t)_{ss^\prime}$ is a distribution obtained applying the Wigner transform to the density matrix, which generalizes the semiclassical phonon population appearing in the Peierls-Boltzmann formalism. The time-evolution of such quantity is regulated by the WTE, which is used to compute the time derivative of $E(\bm R, t)$. The (local) heat flux $\tens{J}(\bm R,t)$ obtained through the continuity equation $\pdv{t}E(\bm R, t) {=} {-}{\grad}\vdot \tens{J}(\bm R,t)$ is related to the expectation value of the total Wigner heat-flux operator $\hat{\tens{J}}$ presented here in Eq.\ \eqref{J WBTE} by $\tfrac{1}{N_c}\sum_{\bm {R}}\tens{J}(\bm R,t) {=} \Tr [\hat{\rho}(t) \hat{\tens{J}}]$ (cf.\ Sec.\ V.A of Ref.\  \cite{long_paper}). This stands also as  proof of  consistency of the definition of the Wigner heat flux, obtained equivalently following the route of WTE as presented in Ref.\  \cite{simoncelli2019unified,long_paper} or Hardy's method introduced in Ref.\  \cite{hardy1963energy} and outlined in Appendix \ref{app:heat flux}.

The main  difference between the two expressions for the heat-flux operator \eqref{J hardy} and \eqref{J WBTE} is the lack in \eqref{J WBTE}, derived from Wigner local energy, of the ``anomalous'' antiresonant term \eqref{J ares}. In addition, 
two different definitions of the velocity matrix are found. The one presented in \eqref{velocity hardy} is the one originally discussed by Hardy and also the most commonly found in literature  \cite{isaeva2019modeling,maradudin1964lattice,semwal1972thermal,allen1993thermal,dangic2020origin}; it consists of matrix elements of the gradient of the dynamical matrix, divided by the geometric mean of the frequencies of the modes involved. The velocity matrix defined in the Wigner formalism involves instead the matrix elements of the square root of the dynamical matrix. It can be shown that the latter satisfies the properties \eqref{vel properties} (satisfied also by Hardy's velocity matrix), and that its diagonal elements are still the usual phonon group velocities $\tens{v}(\bm q)_{ss} {=} \grad_{\bm q} \w(\bm q)_s$.  It can be also shown that the two velocity matrices \eqref{velocity hardy} and \eqref{velocity WBTE} are related by 
\begin{equation}
	\label{hardy 2 wigner}
	\bm v(\bm q)_{ss\p} = \frac{\w(\bm q)_{s}\! +\! \w(\bm q)_{s\p}}{2\sqrt{\w(\bm q)_{s}\w(\bm q)_{s\p} }}  \tens{v}(\bm q)_{ss^\prime}. 
\end{equation}
From these remarks, we understand that the two definitions \eqref{J hardy} and \eqref{J WBTE} of the harmonic heat flux are equal when considering diagonal ($s{=}s\p$) or degenerate ($\w(\bm q)_{s}{=}\w(\bm q)_{s\p}$ for  $s{\neq}s\p$) elements of the velocity matrices,  and they both coincide to the second-quantization form of the heat flux \eqref{J pbte} used in the LBTE formalism. Therefore, the differences between the heat-flux operators under scrutiny are solely in the off-diagonal and non-degenerate ($s{\neq}s\p$ and $\w(\bm q)_{s}{\neq}\w(\bm q)_{s\p}$) terms.

\section{Derivation of thermal conductivity}
\label{sec:conductivity} \label{sec:semiclassical}
In this Section we show how to derive the thermal conductivity using a many-body approach. The expressions of the heat flux in terms of the phonon creation and annihilation operators derived in Section \ref{sec:heat flux} can now be used to compute the thermal conductivity in terms of interacting phonon Green's functions using standard methods of perturbation theory. The general procedure is independent from the choice of the heat flux, even though the final result depends on the structure of the heat-flux operator. 
We first present our expressions in terms of the full phonon spectral function (FSF) for both the Hardy and the Wigner heat-flux operators, and then we describe the expressions we get considering the LSFA. 

To compute the correlation function  at real frequencies we will use, instead of the retarded response function \eqref{chi(t)}, its equivalent in the imaginary-time domain: 
\begin{equation}
	\label{chi(tau)}
	\chi^{\a\b}(\tau) = \langle \mathcal{T}_{\tau} \hat{J}^{\a}(\tau)\hat{J}^{\b}(0) \rangle\:.
\end{equation}
We will then implement the standard many-body formalism of finite-temperature Green's functions  \cite{mahan2013many} to compute the Fourier transform on Eq.\ \eqref{chi(tau)} in Matsubara frequency, and from this by analytical continuation $\chi^{\alpha\beta}(\omega{+}i0^+)$. Given the structure of heat fluxes \eqref{J hardy} and \eqref{J WBTE}, the response function \eqref{chi(tau)} is always expressed as a time-ordered product of four phonon operators. By using the Matsubara-Wick theorem  \cite{matsubara1955new} the response function can be expressed in terms of products of two-particle operators or phonon Green's functions (see Appendix \ref{app:conductivity}):
\begin{equation} \label{g(t)}
	g(\bm q,\tau)_s\! =\! -\langle \mathcal{T}_{\tau} \hat a(\bm q,\tau)_s \hat a^\dagger(\bm q,0)_s  \rangle 
\end{equation}
where the time evolution of the operators is intended in the Heisenberg picture $\dv{ \hat a(\bm q, \tau)_s}{\tau} {=} [\hat{H},\hat a(\bm q, \tau)_s]$. As shown diagrammatically in Fig.\ \ref{fig:dressed_bubble} of Appendix \ref{app:conductivity}, we consider explicitly the so-called ``dressed-bubble'' approximation, 
where only diagrams with two separated dressed Green's functions are included, hence neglecting vertex corrections (the implications of such choice will be discussed at the end of the present section). This approximation makes the analytical continuation straightforward, allowing us at the same time to recover in the LSFA the expression found with the WTE  in Ref.\  \cite{simoncelli2019unified} at the same level of approximation. Now, the rationale behind the decomposition \eqref{J res}-\eqref{J ares} becomes  evident: since the resonant and antiresonant heat-flux operators possess different numbers of creation-annihilation operators, we can neglect mixed terms $\langle \mathcal{T}_{\tau}\hat{\bm J}_R(\tau)\hat{ \bm J}_A(0)\rangle$, assuming, as stated above, that no anomalous averages $\langle \mathcal{T}_{\tau} \hat a(\tau) \hat a(0)\rangle$ are present. Even if such averages cannot be discarded in the most general case, such terms are expected to be negligibly small with respect to the regular ones. This follows from the fact that anomalous averages do not posses a zero-th order term in the perturbative expansion, hence such terms will be always one order higher in the self-energy (defined below) with respect to the normal ones \cite{deo1966calculation,behera1967substitutional,meng2015lattice}. 

The resulting expression for the current-current response function is then given by the product of two phonons Green's functions \eqref{g(t)}. For clarity, let us consider for example the response function  obtained from the Wigner heat-flux operator \eqref{J WBTE}:
\begin{equation}
	\begin{split}
		\chi^{\a\b}(\tau){=}\frac{\hbar^2}{N_c^2\mathcal{V}^2} \!\!\sum_{\bm q, ss\p}\!\!\frac{(\omega(\bm q)_{s}\! +\! \omega(\bm q)_{s\p})^2 }{4} \tenscomp{v}^{\a}(\bm q)_{ss^\prime}  \tenscomp{v}^{\b}(\bm q)_{s\p\! s}  \\ \times g(\bm q, \tau)_{s\p}  g(\bm q,- \tau)_s.         \label{chi(tau) wbte}
	\end{split}
\end{equation}
Here $g(\bm q, \tau)_s$ denotes the ``dressed'' phonon Green's functions, in which self-energy effects due to interactions such as anharmonicity or isotopic disorder are considered (\textit{i.e}.\ intrinsic scattering sources). They can be encoded in the phonon spectral density, defined as:
\begin{equation}
	\label{spectral densities}
	b(\bm q, \w)_s = -\frac{\h}{\pi} \Im g(\bm q,i\w_n)_s \Big|_{i\w_n\to \h\w +i0^+} .
\end{equation}
The advantage of a formalism based on the spectral density is the direct connection with the phonon self-energy. In fact,  the two are related through Eq.\ \eqref{spectral densities} via the relation
\begin{equation}
	\label{spectral density and SE}
	b(\bm q, \w)_{s} = \frac{1}{\pi} \frac{\gamma(\bm q, \w)_s }{ (\omega-\omega(\bm q)_s -\Delta(\bm q, \w)_s)^2 + \gamma(\bm q, \w)_s^2}
\end{equation}
where the phonon self-energy is defined as $\pi(\bm q, \w)_s/\hbar = \Delta(\bm q, \w) {-}i \gamma(\bm q, \w)_s$. Using the latter relation, scattering sources of different nature can be readily included in the calculation of thermal conductivity through the phonon self-energy. For example, in our test cases we consider anharmonicity --- besides intrinsic isotope scattering --- only at lowest perturbative order (3-phonon bubble, \textit{i.e}.\ 3-phonon scattering), but Eq.\ \eqref{spectral density and SE} gives a precise and practical recipe on how to implement higher order anharmonicity such as 4-phonon scattering. This is relevant since it  has been shown how  quartic anharmonicity can change drastically diagonal thermal conductivity in certain materials  \cite{li2018high,xia2020high}, while enhancing coherence effects and interband thermal conductivity in others  \cite{xia2020microscopic}.

The full procedure to carry out the Fourier transform of Eq.\ \eqref{chi(tau) wbte} and to perform the analytical continuation is rather straightforward and it is outlined in Appendix \ref{app:conductivity}. The final expression for the thermal conductivity is a function of the phonon spectral densities \eqref{spectral density and SE}, with a structure depending on the details of the heat-flux operator. In the case of Hardy's heat-flux operator \eqref{J hardy} the thermal conductivity is composed of two terms: 
\begin{widetext}
	\begin{subequations}
		\begin{eqnarray}
			&&\kappa = \kappa_R + \kappa_A \label{k hardy} \,,\\
			\kappa^{\a\b}_R {=} \frac{\hbar^2\pi}{N_c\mathcal{V}T}&&\sum_{\bm q, ss\p}\!\!\frac{(\omega(\bm q)_{s}\! +\! \omega(\bm q)_{s\p}\!)^2\!}{4} v^{\a}\!(\bm q)_{ss\p}  v^{\b}\!(\bm q)_{s\p\! s}\!\!\int\!\! \dd{\w} b(\bm q, \w)_{s\p} b(\bm q, \w)_s \left[-\pdv{n_T(\w)}{\hbar\w}\right], \quad\label{k res}\\
			\kappa^{\a\b}_A{=} \frac{\hbar^2\pi}{N_c\mathcal{V}T}&&\sum_{\bm q, ss\p}\!\!\frac{(\omega(\bm q)_{s}\! -\! \omega(\bm q)_{s\p}\!)^2\!}{4} v^{\a}\!(\bm q)_{ss\p}  v^{\b}\!(\bm q)_{s\p\! s}\!\!\int\!\! \dd{\w} b(\bm q, \w)_{s\p} b(\bm q, -\w)_s \pdv{n_T(\w)}{\hbar\w}\,.\quad \label{k ares}
		\end{eqnarray}
	\end{subequations}
\end{widetext}
The result following from the  Wigner definition of the heat-flux operator \eqref{J WBTE} is instead
\begin{widetext}
	\begin{equation}
		\label{k wbte}
		\kappa^{\a\b} {=} 
		\frac{\hbar^2\pi}{N_c\mathcal{V}T}\!\!\sum_{\bm q, ss\p}\!\! \frac{(\omega(\bm q)_{s}\! +\! \omega(\bm q)_{s\p}\!)^2\!}{4} \tenscomp{v}^{\a}\!(\bm q)_{ss\p}  \tenscomp{v}^{\b}\!(\bm q)_{s\p\! s}\!\!\int\!\! \dd{\w} b(\bm q, \w)_{s\p} b(\bm q, \w)_s \left[-\pdv{n_T(\w)}{\hbar\w}\right].
	\end{equation}
\end{widetext}
The structure of thermal conductivity obtained with the Wigner heat-flux operator is completely analogous to the resonant term obtained in Hardy's formalism; the difference between \eqref{k wbte} and \eqref{k res} resides only in the definition of the velocity matrices (see Eq.\ \eqref{velocity hardy},\eqref{velocity WBTE} and \eqref{hardy 2 wigner}). Therefore, the difference between Eq.\ \eqref{k hardy} and Eq.\ \eqref{k wbte} for the total thermal conductivity is due to the presence of the antiresonant term \eqref{k ares} and due to the differences in the off-diagonal phonon velocities. This clearly mirrors the aforementioned discrepancies in the heat-flux operators. 

Nonetheless, from both starting points, one can derive an expression of thermal conductivity in terms of an integral of the product of two FSFs. This analytical structure provides some physical intuition for interband transport in terms of phonon band mixing. For a perfectly harmonic crystal, the phonon spectral densities are $\delta$-functions at the phonon frequencies $\w(\bm q)_s$, \textit{i.e}.\ $b(\bm q, \w(\bm q)_s) {=} \delta(\w{-}\w(\bm q)_s)$, while interactions have the effect of broadening such shapes and shifting the peaks (cf.\ Eq.\ \eqref{spectral density and SE}). If there is a non-zero overlap between spectral densities of two different modes $(\bm q,s){\neq}(\bm q,s\p)$, the two phonon bands are mixed and can give a finite contribution to interband thermal conductivity.

Now we consider the LSFA for the expressions of the thermal conductivity discussed above. This consists in taking the limit where the probability of finding a vibrational excitation with energy $\omega$, momentum $\bm q$ and band $s$ is mostly  concentrated around the non-interacting excitation energy, \textit{i.e}.\ the harmonic frequency $\omega({\bf q})_s$, and the effect of the interaction is to introduce a finite linewidth to each harmonic phonon mode. This is equivalent to the assumption that the spectral density \eqref{spectral density and SE} has a Lorentzian lineshape, as one obtains by approximating the self-energy with its value at the harmonic phonon frequencies, and neglecting its real part, \textit{i.e}.\ $\gamma(\bm q, \w)_s {\simeq} \gamma(\bm q,\w(\bm q)_s)_s {\to} \gamma(\bm q)_s$ and $\Delta(\bm q, \w)_s {\simeq} 0$. If we further approximate the derivative of the Bose function with its value at the harmonic frequencies, we get that the spectral integrals in \eqref{k res},\eqref{k ares},\eqref{k wbte} consist in a convolution of two Lorentzians, that can be computed analytically. With this procedure, we get from Eq.\ \eqref{k wbte}, obtained using the Wigner heat-flux operator, the following expression for the thermal conductivity in the LSFA
\begin{widetext}
	\begin{equation}
		\label{k wbte rta}
		\kappa^{\a\b}{=}\frac{1}{N_c\mathcal{V}}\!\sum_{\bm q, ss\p}\!\!\frac{\w(\bm q)_{s}\! +\! \w(\bm q)_{s\p}\!}{4} \tenscomp{v}^{\a}\!(\bm q)_{ss\p}  \!\tenscomp{v}^{\b}\!(\bm q)_{s\p \!s}\left[\frac{C(\bm q)_s}{\w(\bm q)_s} + \frac{C(\bm q)_{s\p}}{\w(\bm q)_{s\p}}\right] \frac{\frac{1}{2}(\Gamma(\bm q)_s\! + \Gamma(\bm q)_{s\p} )\!}{ (\w(\bm q)_s\! -\! \w(\bm q)_{s\p}\! )^2\!+\! \tfrac{1}{4}(\Gamma(\bm q)_s\! +\! \Gamma(\bm q)_{s\p}\!)^2} 
	\end{equation}    
\end{widetext}
where the modal heat capacity  $C(\bm q)_s $ has been defined in \eqref{k pbte}, and we have used the relation $\gamma(\bm q)_s {=} \tfrac{1}{2}\Gamma(\bm q)_s $ between the scattering rate $\Gamma(\bm q)_s$ used in Boltzmann SMA (\textit{i.e}.\ the linewidth) and the phonon self-energy $\gamma(\bm q)_s$ (see e.g.\ Ref.\  \cite{mahan2013many}, Sec.\ 8.1). This expression coincides exactly with the SMA of WTE obtained in Ref.\  \cite{simoncelli2019unified}, as further discussed below.
Using the same procedure for the thermal conductivity \eqref{k hardy} obtained from Hardy heat flux we get
\begin{widetext}
	\begin{subequations}
		\begin{eqnarray}
			&&\kappa {=} \kappa_R + \kappa_A \, , \label{k hardy rta} \\
			\kappa^{\a\b}_R {=} &&\frac{1}{N_c\mathcal{V}}\!\sum_{\bm q, ss\p}\frac{\w(\bm q)_{s}\! +\! \w(\bm q)_{s\p}\!}{4} v^{\a}\!(\bm q)_{ss\p}  v^{\b}\!(\bm q)_{s\p\! s} \left[\frac{C(\bm q)_s}{\w(\bm q)_s}\! +\! \frac{C(\bm q)_{s\p}}{\w(\bm q)_{s\p}}\right]\frac{\frac{1}{2}(\Gamma(\bm q)_s\! + \Gamma(\bm q)_{s\p} )\!}{ (\w(\bm q)_s\! -\! \w(\bm q)_{s\p}\! )^2\!+\! \tfrac{1}{4}(\Gamma(\bm q)_s\! +\! \Gamma(\bm q)_{s\p}\!)^2} \,,\label{k res rta} \\
			\kappa^{\a\b}_A {=} &&\frac{1}{N_c\mathcal{V}}\!\sum_{\bm q, ss\p}\frac{\w(\bm q)_{s}\! -\! \w(\bm q)_{s\p}\!}{4} v^{\a}\!(\bm q)_{ss\p}  v^{\b}\!(\bm q)_{s\p\! s} \left[\frac{C(\bm q)_{s\p}}{\w(\bm q)_{s\p}}\! -\! \frac{C(\bm q)_{s}}{\w(\bm q)_{s}}\right] \frac{\frac{1}{2}(\Gamma(\bm q)_s\! + \Gamma(\bm q)_{s\p} )\!}{ (\w(\bm q)_s\! +\! \w(\bm q)_{s\p}\! )^2\!+\! \tfrac{1}{4}(\Gamma(\bm q)_s\! +\! \Gamma(\bm q)_{s\p}\!)^2} \,.\label{k ares rta}
		\end{eqnarray}
	\end{subequations}
\end{widetext}
The latter expressions elucidate the nomenclature for Hardy's heat-flux operator: the amplitude of the resonant thermal conductivity \eqref{k res rta} is maximum when the frequencies of the two modes resonate, \textit{i.e}.\ when $\w(\bm q)_{s}{\simeq}\w(\bm q)_{s\p}$, whereas the antiresonant \eqref{k ares rta} term is maximum when $\w(\bm q)_{s}{\simeq} {-} \w(\bm q)_{s\p}$. Since the phonon frequencies are positive definite, the condition that maximizes the antiresonant terms cannot be met, hence such terms are expected to be much smaller than the resonant ones. Once again, the resonant thermal conductivity \eqref{k res rta} resulting from Hardy's heat flux and the one derived from Wigner \eqref{k wbte rta} share a common structure and they differ only in the {non-degenerate} interband $s{\neq}s\p$ terms, due to the already discussed difference in the definition of the velocity matrices \eqref{hardy 2 wigner}.
\begin{figure}[h]
	\centering
	\includegraphics[width=0.5\textwidth,keepaspectratio]{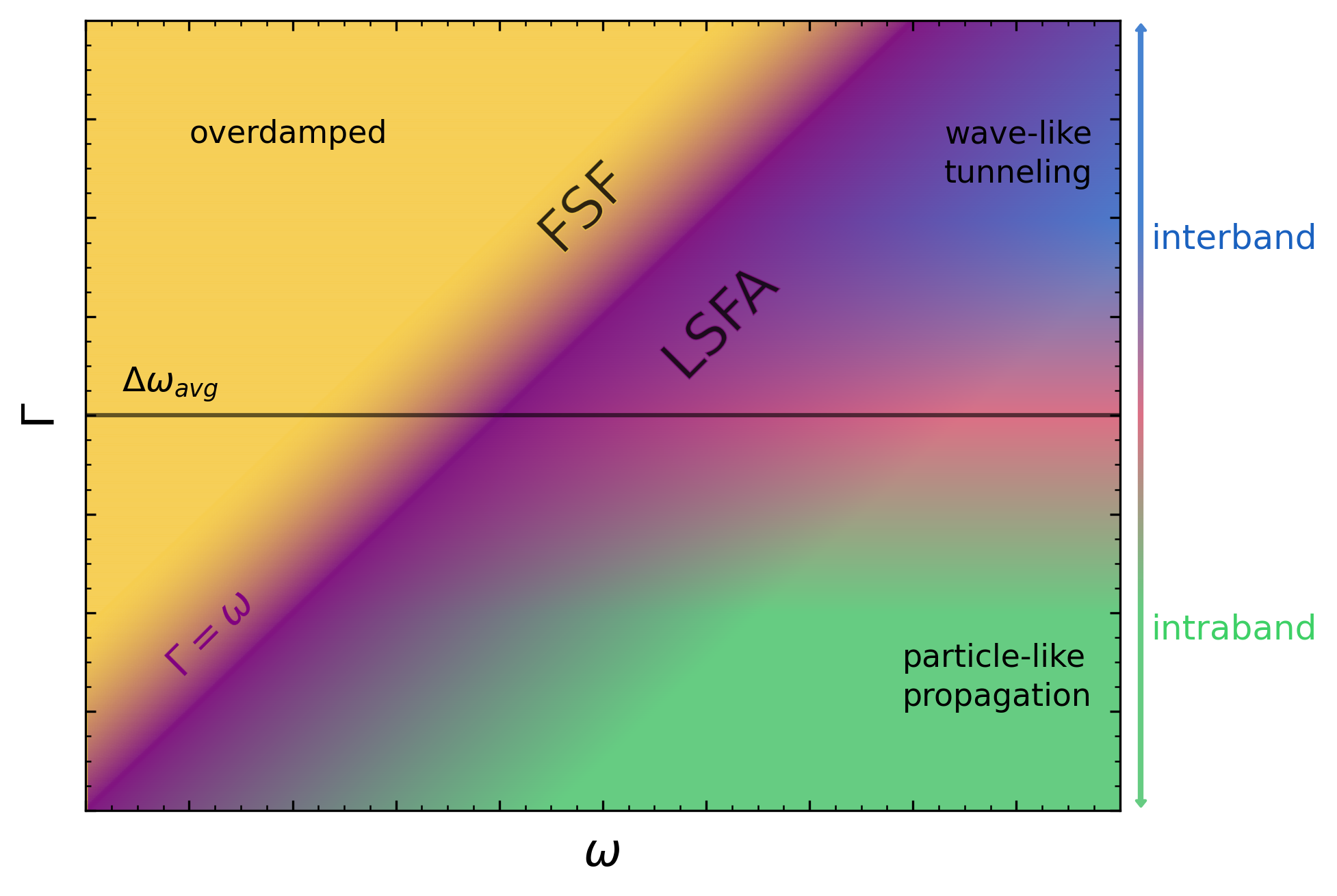}
	\caption{Diagram of the thermal transport regimes considering phonon linewidths ($\Gamma$) as a function of frequencies ($\omega$). When the linewidths are larger than the frequencies, phonons are ill-defined and the vibrational modes are overdamped (gold area). In the latter regime, vibrational excitation must be discussed using full phonon spectral functions (FSF). When the linewidths are smaller than the frequencies (purple area), the Lorentzian spectral function approximation (LSFA) becomes feasible. The strength of phonon damping determines the transition from the particle-like regime (green area, where heat is mainly carried by phonon wavepackets that behave as particles of a gas), to the wave-like tunneling regime (blue area, where the effective overlap between phonon states allows them to interfere and enables interband tunneling transport).
		The transition from the particle-like to the wave-like regime is non-sharp (red shaded area), and phonons having a linewidth around the average interband spacing $\Delta \w_{\text{avg}}$ (horizontal line) are at the center of this non-sharp crossover, behaving both particle- and wave-like. {Both wave-like and particle-like conduction are well described in the LSFA, provided that heat hydrodynamics is negligible (see text).}  }
	\label{fig:phase_diag}
\end{figure}

In order to discuss the differences between the expressions of thermal conductivity derived so far, it is instructive to consider different regimes for thermal transport. The two relevant quantities to be considered are the relative strength of the phonon linewidth $\Gamma(\bm q)_s$ with respect to the harmonic frequencies $\w(\bm q)_s$, and the magnitude of  $\Gamma(\bm q)_s$  with respect to the typical spacing among phonon branches, that can be roughly estimated as $\Delta \w_{\text{avg}}{=}\frac{\w_{\text{max}} }{ 3N_{at} } $ ($\w_{\text{max}}$ is the maximum phonon frequency and $3N_{at}$ the number of phonon bands), see Fig.\ \ref{fig:phase_diag}. In general, whenever $\Gamma(\bm q)_s{\ll}  \w(\bm q)_s$ phonon states are well defined and one can safely approximate the phonon spectral function with a Lorentzian: this is the regime where the LFSA holds  \cite{lifshitz2013statistical} (purple shaded section of Fig.\ \ref{fig:phase_diag}). If $\Gamma(\bm q)_s$ is also lower than $\Delta \w_{\text{avg}}$, phonon branches do not overlap significantly and the Peierls-Boltzmann semiclassical theory becomes accurate (green shaded area in Fig.\ \ref{fig:phase_diag}): phonons wavepackets propagate as particle of a gas, with well defined momentum $\bm q$ (the central momentum of the wavepacket), energy $\hbar\omega(\bm q)_s$, and scattering rate $\tau(\bm q)_s{=}\tfrac{1}{\Gamma(\bm q)_s}$. As the linewidth increases and overcomes  $\Delta\w_{\text{avg}}$, vibrational modes start to get mixed. In this regime, the wave-like nature of the atomic vibrations emerges, and heat conduction can also occur through the tunneling of phonons between overlapping bands. Note that there is no sharp separation between the two regimes: a phonon mode can contribute to thermal transport both in a particle-like and wave-like fashion, hence participating equally in intraband and interband conduction  \cite{simoncelli2019unified,long_paper} (red shaded area of Fig.\ \ref{fig:phase_diag}). This threshold has been denoted in Ref.\  \cite{long_paper} as the Wigner limit, 
and it has been shown that phonons around this threshold contribute comparably to the particle-like (intraband) and wave-like (interband) conductivity. On the other hand, the system enters a completely different regime when $\Gamma(\bm q)_s{\gtrsim}\w(\bm q)_s$ (gold-colored side of Fig.\ \ref{fig:phase_diag}), since in this case the LSFA breaks down and the quasiparticle picture becomes invalid, with phonons losing their ``identity'' as heat carriers due to the strong interactions. In this regime the FSF must be used to compute thermal conductivity, considering the frequency dependence of the phonon self-energy. Also, in this case, the crossover to the overdamped regime is not defined by a sharp transition.

To summarize, while the expressions of thermal conductivity derived in this section in terms of the FSFs are valid throughout all the transport regimes regardless of phonon damping (within the limits of validity of perturbation theory), the expressions derived in the LSFA are correct only far from the overdamped regime. Nonetheless, even in the LSFA, the expression for the thermal conductivity we use consists of an extension of the BTE considering the interband contribution to thermal transport.

However, we should mention that some care is needed in considering a general regime of validity of the dressed-bubble approximation. In fact, there is a transport regime where a theoretical justification of such approximation should not hold. This is the so-called hydrodynamical regime of thermal transport \cite{guyer1966thermal}, where momentum-conserving (normal) dominates over the resistive (Umklapp) scattering \cite{cepellotti2015phonon}. 
	The Umklapp scatterings are 3-phonon scattering events where crystal momentum is not conserved \cite{ziman2001electrons,peierls1955quantum}.
	The normal phonon scatterings are instead scattering events that conserve crystal momentum. In practice, Umklapp scattering limits heat conduction, while normal scattering do not degrade the heat flux but redistribute it across different phonon modes \cite{fugallo2014thermal,cepellotti2015phonon}.
	In a many-body language, it is known that in order  to account for the fact that normal scatterings do not degrade the flux, one has to consider vertex corrections in the current-current response function, that are instead automatically included in the full solution of the BTE \cite{sun2010lattice,mahan2013many,fugallo2014thermal,cepellotti2015phonon,physrevx.10.011019}. Neglecting such effects consists for the intraband conductivity (in the LSFA) in approximating the so-called transport relaxation time with the quasiparticle relaxation time (phonon self-energy). 
	The validity of such an approximation scheme can be verified \textit{a posteriori} by comparing the results for the thermal conductivity obtained with a LBTE approach obtained from a full diagonalization procedure with the ones obtained with SMA. However, it is generally understood \cite{lindsay2016first,simoncelli2019unified,long_paper} that heat hydrodynamics appears typically in materials with large values of thermal conductivity (${\gtrsim}10^3$ W/mK around room temperature for typical carbon-based materials \cite{ward2009ab,fugallo2014thermal,lindsay2016first}), in which it has been shown that accounting for phononic collective excitations with a full diagonalization procedure of LBTE overcomes the failure of the SMA in describing simple crystals \cite{cepellotti2016thermal,physrevx.10.011019}. Instead, for materials with ultralow thermal conductivity, the results obtained from the solution of the LBTE determined within the SMA approximation are practically indistinguishable from the exact solution of the LBTE \cite{long_paper}. Since the main interest of the present work is to investigate the physics of interband thermal transport in ultralow-conductivity complex materials in temperature ranges where Umklapp scatterings are predominant, we do not expect to find any signature of heat hydrodynamics (\textit{i.e}.\ vertex corrections effects) in our test cases and our approximation is expected to yield quantitatively reliable results. Anyhow, we have performed the abovementioned tests comparing our result for intraband thermal conductivity obtained via the dressed-bubble LSFA with the one obtained with a (computationally much more demanding) full-diagonalization procedure of the LBTE, finding no quantitatively relevant numerical differences for the materials under scrutiny.

In general, the LBTE and the WTE are able to describe the physics of intraband transport accounting also for the hydrodynamic regime of thermal transport, a feature missing in this work. On the other hand, interband transport is not taken into account by the LBTE and is described by the WTE only far from the overdamped regime, while here interband transport is treated in full generality in all the regimes including the overdamped one.

\subsection{Comparison with other approaches  } \label{subsec:comparison}
In this Section, we will discuss differences and analogies between the present expressions of thermal conductivity derived from the Hardy or Wigner heat fluxes with respect to the other results in the literature. Let us start by considering the comparison with the results derived using Hardy heat flux \eqref{J hardy}, as used both in older theoretical studies (Ref.\  \cite{maradudin1962scattering,semwal1972thermal}) and more recently in Ref.\  \cite{pereverzev2018theoretical,isaeva2019modeling}. All these rely on a Green-Kubo approach, analogous to the one employed here, but with some differences that will be discussed below.

First of all, the present definition \eqref{g(t)} of the phonon Green's function $g(\bm q,\tau)_s$ is different from the one commonly found in  literature  \cite{mahan2013many,abrikosov2012methods}, corresponding to $G(\bm q, \tau)_s{=} {-}\langle \mathcal{T}_{\tau}[ \hat a(\bm q, \tau)_s  {+} \hat a^\dagger({-}\bm q,\tau)_s ] [\hat a({-}\bm q,0)_s  {+} \hat a^\dagger(\bm q, 0)_s ]  \rangle$; the two are related simply by $G(\bm q, \tau)_s= g(\bm q, \tau)_s {+}g(\bm q,{-}\tau)_s  $   (further discussion in Appendix \ref{app:greens functions}). The standard definition can be motivated by the fact that interaction terms for phonons depend only on the atomic displacement $\hat{u}{\sim} \hat a{+}\hat a^\dagger$. On the other hand, in the definition of the heat flux  the momentum $\hat{p}$ of the atoms appears explicitly (cf.\ Eq.\ \eqref{J = up}). Since $\hat{u}$ and $\hat{p}$ are independent variables, we cannot express the correlations for the thermal current only in terms of the Green's function $G(\tau){\sim} \langle\hat{u}(\tau)\hat{u}(0)\rangle$, but one must also consider an additional mixed Green's function such as $\tilde{G}(\tau){\sim}\langle\hat{p}(\tau)\hat{u}(0)\rangle{\sim} g( \tau) {-}g({-}\tau) $  \cite{maradudin1964lattice,semwal1972thermal}.

The expressions \eqref{k hardy}-\eqref{k ares} for the thermal conductivity derived from Hardy's heat flux \eqref{J hardy} in terms of the FSF coincide with the results first proposed in Ref.\  \cite{maradudin1964lattice} and Ref.\  \cite{semwal1972thermal}. In Ref.\  \cite{semwal1972thermal} the thermal conductivity is obtained using a double-time Green's function approach  \cite{semwal1972thermal,pathak1965theory}, employing a particular decoupling scheme --- cf.\ Eq.\ \eqref{decoupling} ---  to the heat-flux autocorrelation function (two-particle Green's function) directly in the time domain. We will show in Appendix \ref{app:greens functions} how this decoupling scheme is equivalent to the present dressed-bubble approximation. Given such an equivalence,  the present framework based on a diagrammatic expansion provides a transparent identification of the approximation used, and also shows how to systematically refine the results obtained by accounting for higher-order processes.  The approach followed by Ref.\  \cite{maradudin1964lattice} is more similar to the one presented here, but the approximation of the heat-flux autocorrelation is considered only for isotope scattering and without any diagrammatic interpretation. Both Ref.\  \cite{maradudin1964lattice} and Ref.\  \cite{semwal1972thermal} rely on a formalism based on phonon displacement operators $\hat A(\bm q)_s {=}\hat a(\bm q)_s  {+} \hat a^\dagger({-}\bm q)_s$, which leads to the definition of 4 kind of Green's functions (2 of which are independent) \cite{maradudin1964lattice,semwal1972thermal}. Instead, here we only consider phonon operators $\hat a(\bm q)_s$ and, as mentioned above, only one type of Green's function is needed, which greatly simplifies the notation. In this regard, we argue that the present formalism is more compact and consists of a more suitable choice for the treatment of lattice thermal transport in a many-body theoretical approach. Besides mathematical ease, the thermal conductivity \eqref{k hardy}-\eqref{k ares} with this choice of the phonon Green's function features separate resonant and antiresonant contributions, whereas in the result proposed in both Ref.\  \cite{maradudin1964lattice} and Ref.\  \cite{semwal1972thermal} those terms are mixed (see Appendix \ref{app:greens functions} for details). As we shall see, the antiresonant terms are typically negligible with respect to the resonant ones, hence our expression \eqref{k hardy}-\eqref{k ares} is more effective in highlighting leading order terms. 

\begin{figure*}[t]
	\centering
	\includegraphics[width = \textwidth,keepaspectratio]{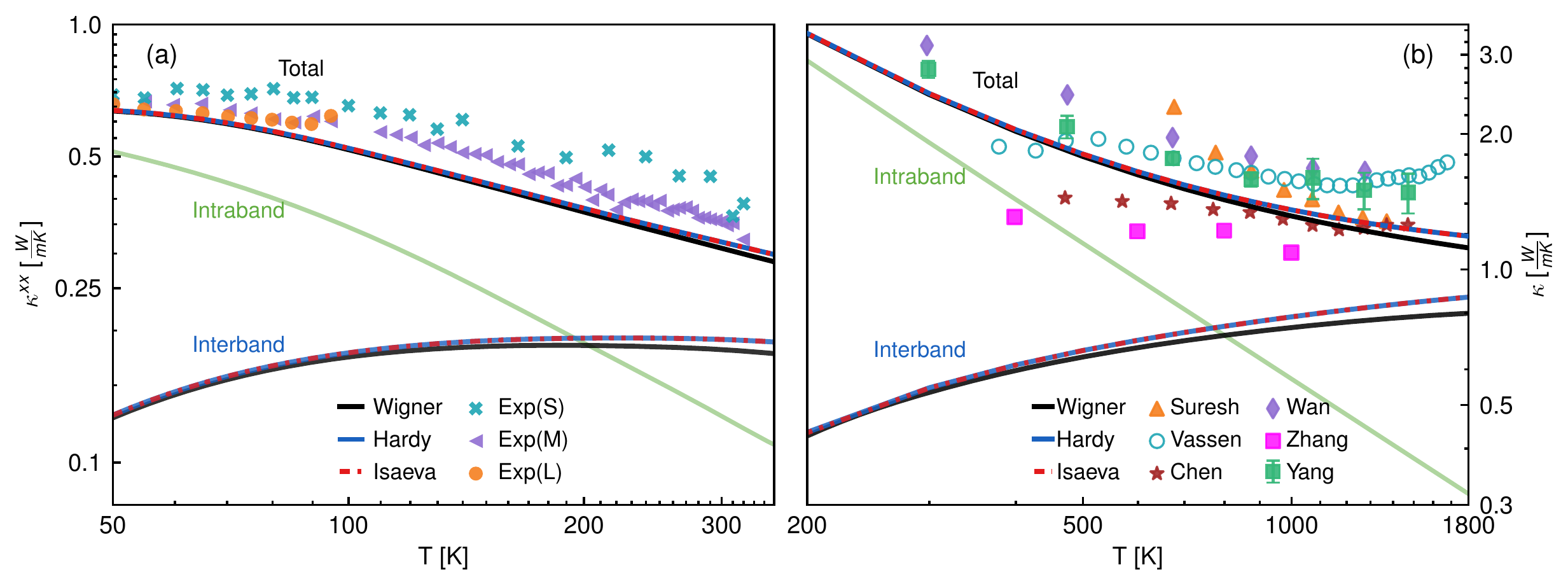}
	\caption{
		Lattice thermal conductivity  of \ch{CsPbBr_3} (a) and \ch{La2Zr2O7} (b). Experimental data for \ch{CsPbBr3} are taken from Ref.\  \cite{wang2018cation} and ``Exp(S)”, “Exp(M)” and “Exp(L)” refer to measurements  of $\kappa^{xx}$ in nanowires of increasing section from smallest (S) to largest (L). 
		Experimental data for \ch{La2Zr2O7} are taken from Suresh \textit{et al}.\  \cite{suresh1997investigation}, Vassen \textit{et al}.\  \cite{vassen2000zirconates}, Chen \textit{et al}.\ \cite{chen2009thermophysical}, Wan \textit{et al}.\  \cite{wan2010glass}, Zhang \textit{et al}.\ \cite{zhang2020microstructure}, Yang \textit{et al}.\  \cite{yang2016effective}. Lines in different colors represents theoretical prediction from different formulations of  thermal conductivity in the LSFA. For \ch{CsPbBr3} we report the conductivity tensor component $\kappa^{xx}$, for \ch{La2Zr2O7} we report a component of the isotropic conductivity tensor. 
		Black, result from Wigner heat-flux from Eq.\ \eqref{k wbte rta}. Blue, result from Hardy heat-flux from Eq.\ \eqref{k hardy rta}-\eqref{k ares rta}. Red, result obtained following the approach of Ref.\  \cite{isaeva2019modeling} (Isaeva \textit{\textit{et al}}.\ \!, reported in the Appendix in Eq.\ \eqref{k isaeva}). In the lower part of the figures  the  interband contributions to thermal conductivity {are reported}, {which are} obtained by considering only {the off-diagonal and non-degenerate terms} ($s{\neq} s\p$ with $\omega(\bm{q})_s{\neq} \omega(\bm{q})_{s'}$)  in the corresponding equations. {The intraband term is reported in light green, it is} obtained by {considering only the diagonal ($s{=}s\p$) or degenerate ($s{\neq}s\p$ with $\omega(\bm{q})_s{=}\omega(\bm{q})_{s'}$) terms }  in the corresponding equations {and in its the same in all the approaches considered}. The results  for the total conductivity (upper lines) are obtained by summing the corresponding  interband term (lower lines) with the intraband part (light green).        }
	\label{fig:kappa_exp}
\end{figure*}

We can compare the present result for the thermal conductivity in the LSFA derived from the Hardy heat-flux operator \eqref{k hardy rta}-\eqref{k ares rta} with the recent results of Refs.\  \cite{isaeva2019modeling,pereverzev2018theoretical}, obtained with a Green-Kubo approach where a constant scattering rate is assigned to each phonon mode. We note that the results for the thermal conductivity presented in Refs.\  \cite{isaeva2019modeling,pereverzev2018theoretical} slightly differ from \eqref{k hardy rta}-\eqref{k ares rta}, in the fact that the interband conductivity of Refs.  \cite{isaeva2019modeling,pereverzev2018theoretical} features finite differences of the Bose-Einstein functions instead of the modal specific heats prefactors (related to the Bose-Einstein derivatives) featuring in \eqref{k hardy rta}-\eqref{k ares rta}. We give a possible explanation of this discrepancy --- along with some more details on the comparison --- in Appendix \ref{app:greens functions}, where we argue that such differences stem from a subtlety in the derivation of thermal conductivity when considering phonon damping directly in the time-domain or in the frequency-domain expression of the phonon Green's functions. To our present knowledge, we find that this difference has only a formal valence since consistency is found in the numerical calculation of thermal conductivity for both test case materials considered in this work, as shown in Sec.\ \ref{subsec:applications}.

Concerning the results obtained with Wigner heat flux \eqref{J WBTE}, we emphasize that Eq.\ \eqref{k wbte} represents a refinement of the WTE thermal conductivity formula in the SMA approximation  \cite{simoncelli2019unified,long_paper} and is one of the main results of this work. In fact, besides reducing in the LSFA limit to the SMA of the WTE, it is valid in general for every shape of the spectral density, \textit{i.e}.\ for every structure of the phonon self-energy. Therefore, the expression for interband thermal conductivity derived here can also be used in the overdamped regime where the spectral densities depart from the Lorentzian lineshape, e.g.\ near the edge of a structural phase transition  \cite{aseginolaza2019phonon,lanigan2021two}. On the other hand, the WTE result presented in  \cite{simoncelli2019unified} has the advantage of considering fully the LBTE result for the intraband conduction, hence being able to describe the hydrodynamic transport regime that we miss in the dressed-bubble approximation  \cite{mahan2013many}, as stressed in the discussion of Fig.\ \ref{fig:phase_diag}.

As mentioned, Eq.\ \eqref{k wbte rta} derived as the LSFA of Eq.\ \eqref{k wbte} coincides exactly with the SMA of the WTE obtained in Ref.\  \cite{simoncelli2019unified}. The result \eqref{k wbte rta} proves how the same expression for thermal conductivity can be achieved using the Wigner transport equation or a Green-Kubo approach, thus consolidating the theoretical framework of the Wigner formulation of thermal transport while also clarifying how the results for thermal conductivity do not depend on the theoretical procedure employed to derive them.

Finally, we stress that {when heat hydrodynamics can be neglected}, all the formulations of thermal conductivity previously derived admit an intraband contribution which reduces in the semiclassical limit --- \textit{i.e}.\  neglecting interband terms in the LSFA ---  to the result from Peierls-Boltzmann transport equation in SMA  \cite{ziman2001electrons,srivastava2019physics}. In fact, if we take $s{=}s\p$ both in \eqref{k wbte rta} and in \eqref{k hardy rta}-\eqref{k ares rta}  we get exactly the result for the BTE presented in Eq.\ \eqref{k pbte}.

\begin{figure*}[t]
	\centering
	\includegraphics[width = 0.98\textwidth,keepaspectratio]{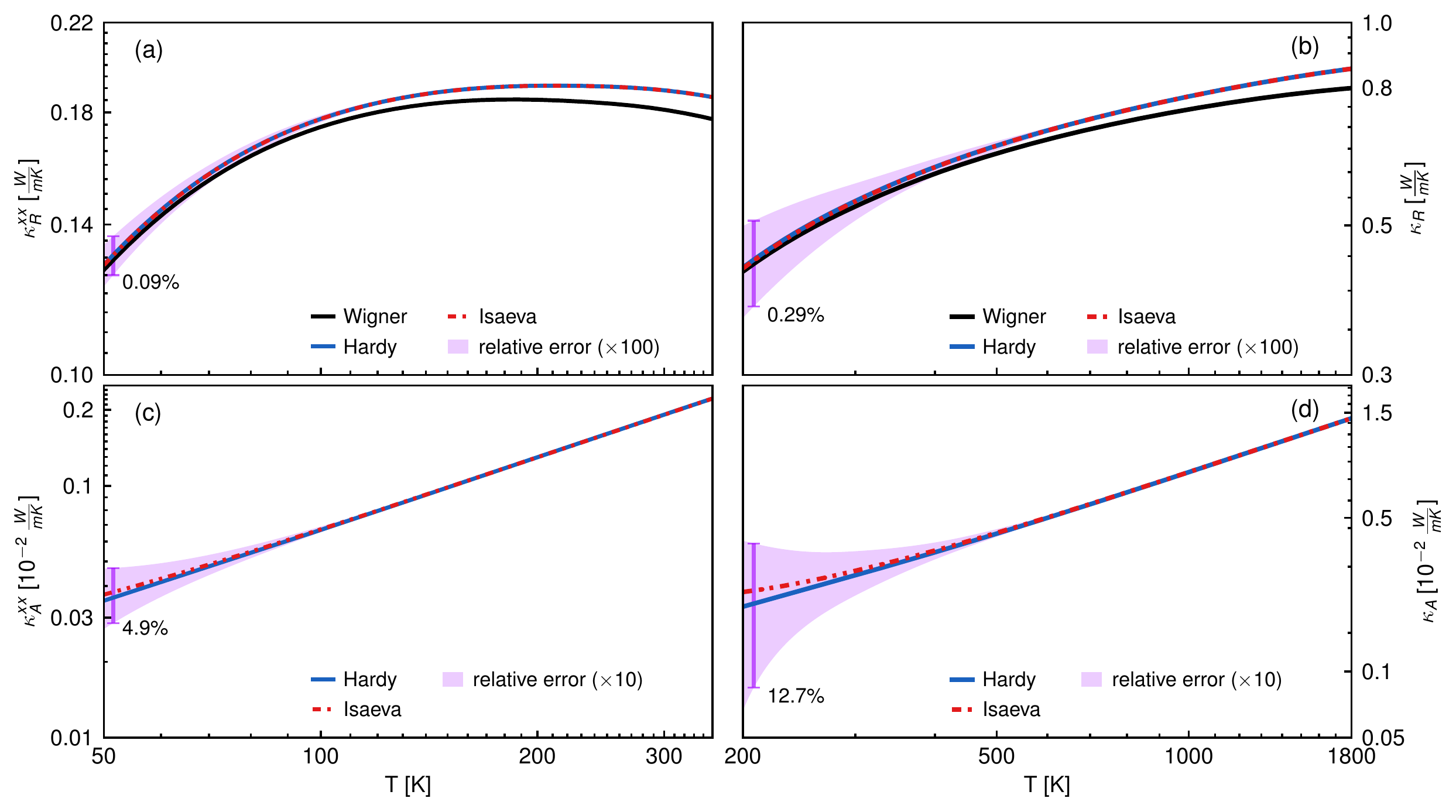}
	\caption{Comparison between the interband thermal conductivity obtained by means of the Hardy (blue, Eq.\ \eqref{k res rta}-\eqref{k ares rta}) heat flux, the Wigner heat flux (black, Eq.\ \eqref{k wbte rta}) and the Hardy heat flux in Ref.\  \cite{isaeva2019modeling} (red, Isaeva \textit{et al}., reported in the Appendix in Eq.\ \eqref{k isaeva}). Left column: results for \ch{CsPbBr3} (a-c). Right column: results for \ch{La2Zr2O7} (b-d). For the case of the Hardy heat flux, we further distinguish between the resonant contribution (upper panels, a, b) and the antiresonant one (lower panels, c, d). The different scales in the latter plots elucidate how the antiresonant thermal conductivity is typically two orders of magnitude smaller than the resonant one, and it is mostly negligible. We also notice that in the LSFA the quantitative differences between the present results \eqref{k res rta}-\eqref{k ares rta} and that those of Ref.\  \cite{isaeva2019modeling} for the Hardy heat flux are numerically negligible. This is further evidenced by the   
		pink shading, which represents the relative error between the blue and red lines, rescaled by a multiplicative factor for proper visualization. A single value of the relative error is reported as a percentage to set the scale.    }
	\label{fig:kappa_theo}
\end{figure*}

\subsection{Application to test cases \ch{CsPbBr3} and \ch{La2Zr2O7}}
\label{subsec:applications}
We will now apply the result of thermal conductivity discussed and derived in the previous Section to two test case materials: \ch{CsPbBr3} and \ch{La2Zr2O7}. \ch{CsPbBr3} belongs to the family of lead-halide perovskite, interesting for their ultralow thermal conductivity and potential candidates for thermoelectric energy conversion  \cite{lee2017ultralow,pisoni2014ultra,wang2018cation}. \ch{La2Zr2O7} is a pyrochlore bulk insulator characterized by thermal stability and ultralow thermal conductivity, thus it is of great interest for thermal barrier coating applications  \cite{zhang2017lanthanum}. It is also an important test case for the calculations since interband lattice heat transport becomes predominant in this material at high temperatures (namely in the range that concerns applications). For the details of the vibrational properties of these materials such as phonon bands  and linewidths, we refer the reader to Ref.\ \cite{simoncelli2019unified} and Ref.\ \cite{long_paper} respectively for \ch{CsPbBr3} and \ch{La2Zr2O7}.  

For these systems we will implement the formulas for thermal conductivity obtained in the LSFA, namely \eqref{k hardy rta} from the Hardy heat flux and \eqref{k wbte rta} for the Wigner heat flux (that coincides with that of Ref.\  \cite{simoncelli2019unified}). To enrich the comparison outlined in Sec.\ \ref{subsec:comparison}  between the present and previous results, we also implement the expression of the thermal conductivity presented in Ref.\  \cite{isaeva2019modeling}, derived using Hardy heat flux (the explicit expression is reported in the Appendix as Eq.\ \eqref{k isaeva}). We did not implement the expression of the thermal conductivity proposed in Ref.\  \cite{pereverzev2018theoretical} since it was derived using yet another definition for the heat-flux operator, which consists of a slight modification of Hardy heat flux \eqref{J hardy}, even though some plausible arguments on why the result for the thermal conductivity proposed in Ref.\  \cite{pereverzev2018theoretical} should be similar to the one proposed in Ref.\  \cite{isaeva2019modeling} for these systems are sketched in Appendix \ref{app:greens functions}.  The phonon scattering rates are computed considering the lowest-order 3-phonon scattering and the intrinsic isotopic scattering. Details of the numerical calculation can be found in Ref.\  \cite{long_paper}.

The application of the LSFA to these test cases is motivated by the fact that in the temperature range considered the vast majority of vibrational modes of both \ch{La2Zr2O7} and \ch{CsPbBr3} feature linewidths well below the value of their harmonic frequencies (cf.\  Fig.8 of Ref.\  \cite{long_paper}); hence we can treat both materials as safely far from the overdamped regime. It is worth noting that for the perovskite \ch{CsPbBr3} a structural phase transition from the orthorhombic to the tetragonal phase is predicted for $T_c^{ot}{\sim} 361\,K$, and another from tetragonal to cubic at $T_c^{tc}{\sim} 403 \,K$   \cite{hirotsu1974structural}. It has been shown recently in Ref.\  \cite{lanigan2021two} how these two phases are characterized by extreme phonon damping, as evidenced by inelastic neutron scattering data signaling very broad phonon spectral profiles. The latter study testifies how even if the LSFA may hold for the orthorhombic phase, theoretical prediction for thermal conductivity of \ch{CsPbBr3} at higher temperatures must be done with a formalism capable of describing overdamped phonon modes, considering the FSFs as in the present results of Eqs.\ \eqref{k hardy}-\eqref{k ares} and \eqref{k wbte}.

The results are shown in Fig.\ \ref{fig:kappa_exp}. Here one sees that the different formulations for thermal conductivity give basically the same numerical results for the total thermal conductivity in both systems. In addition, the values for the thermal conductivity obtained are in good agreement with experimental data. As already mentioned, all the formulations studied lead to the same expression for the diagonal Peierls-Boltzmann conductivity, reported in light green and denoted as ``intraband'' in Fig.\ \ref{fig:kappa_exp}. This contribution, following the expected typical $T^{{-}1}\,$ decay  \cite{peierls1929kinetischen}, is however in broad disagreement with the experimental data. Such a disagreement is corrected by the interband terms, reported in the lower side of Fig.\ \ref{fig:kappa_exp}. The results presented in Fig.\ \ref{fig:kappa_exp} show how the interband terms provide a sizeable contribution to the thermal conductivity especially at high temperatures, highlighting the importance of a correct description of interband transport in systems of this kind. 

On the other hand, very small numerical differences are found between the Hardy and Wigner formulations of the heat flux. This is better seen in Fig.\ \ref{fig:kappa_theo}, where we provide a closer look at the interband contribution to the thermal conductivity as resulting from different formulations of the heat flux. In particular, the two results obtained starting from the Hardy heat flux are consistent with each other, with small discrepancies with the Wigner result when the temperature is increased. This is most likely a consequence of the fact that in these materials the phonon bands that build up most of the interband thermal conductivity are close in energy, \textit{i.e}.\  $\w(\bm q)_s {\simeq} \w(\bm q)_{s\p}$ in \eqref{k hardy rta}. If this is the case, the differences between the velocities \eqref{hardy 2 wigner} are small, and the finite differences of the Bose functions well approximate the modal heat capacities which appear in our formulation. The antiresonant terms in Fig.\ \ref{fig:kappa_theo}(c-d) consist in less than $1\%$ of total thermal conductivity through all the temperature ranges and are thus mostly negligible. The relative differences with the results from Ref  \cite{isaeva2019modeling} decrease when increasing temperature, especially for the antiresonant term. This can be understood from the fact that when the temperature is higher than the Debye temperature $\theta_D$, \eqref{k hardy rta}-\eqref{k ares rta} and  Ref.\  \cite{isaeva2019modeling}  are the same, since for $T{\gg}\theta_D$ it holds $n(\omega(\bm q)_s){\simeq} k_BT/\hbar\w(\bm q)_s \: \forall (\bm q, s)$ .

\section{Heat flux and gauge invariance }
\label{sec:discussion}
In this Section, we comment on the numerical results of the thermal conductivity obtained from the different definitions of the heat flux in the LSFA presented in Sec.\ \ref{subsec:applications}, discussing whether the agreement between the computed thermal conductivities could have been foreseen as a consequence of some broader invariance principle of thermal transport, that ensures consistency of transport coefficients regardless of the microscopic details. In fact, the real physical question triggered by the numerical correspondence between the results obtained with the Hardy and the Wigner heat flux concerns the role played by the freedom to partition the total energy in local contributions. In other words, do we have a guiding principle to decide what is the best choice of local energy to compute the thermal conductivity at a certain level of approximation? 

As mentioned in the introduction, a possible way to reconcile results obtained with different definitions of heat flux has been outlined in Ref.\   \cite{ercole2016gauge}. Here the authors show that whenever two choices of local energies differ by the divergence of a bounded vector field, the corresponding heat fluxes lead to the same thermal conductivity. The reason is that the corresponding heat fluxes, obtained from the continuity equation \eqref{continuity1}, differ by a total time-derivative term, which gives no contribution to thermal transport. Such a result, demonstrated in Ref.\  \cite{ercole2016gauge} by using a classical limit for the general formula \eqref{kubo lambda}, can be extended to the quantum case, as shown in Appendix \ref{app:gauge}. The basic mechanism can be easily understood from Eq.\ \eqref{k = chi}: since the thermal conductivity is defined as the $\omega{\rightarrow} 0$ limit of the current-current response function divided by $\omega$, adding to the current a time derivative brings an additional factor $\omega^2$ in the numerator, leading to a vanishing contribution in the zero-frequency limit. Such a result can be used for example to justify why we carried out the coarse-graining procedure outlined in Sec.\ \ref{sec:heat flux} for the Hardy local energy using the equilibrium position $\bm R {+}\bm{\tau}_b$ instead that the instantaneous $\bm R {+}\bm{\tau}_b{+}{\bm u}(\bm R)_{b}$ atomic positions into the $\Delta_\ell$ functions Eq.\ \eqref{hardy h(x)}. Indeed, adding the local displacements ${\bm u}(\bm R)_b$ is equivalent to adding to the heat flux a total time derivative. 

While very attractive and powerful, such gauge invariance does not answer the question posed. First, it provides a sufficient but not necessary condition for the equivalence between two different definitions of the heat flux. In other words, to the best of our knowledge, it has not been proven that all possible partitioning of the total energy must differ by a total derivative. More specifically, we could not prove that the two energy density operators derived from Hardy \eqref{hardy h(R)} and from Wigner \eqref{e(R) WBTE} differ for the divergence of a bounded vector field. Thus, we cannot state that the two quantum heat-flux operators \eqref{J hardy} and \eqref{J WBTE} differ by a total time derivative. 

Secondly, the gauge-invariance principle is valid if the thermal conductivity is derived from the exact current-current correlation function, which is the one typically obtained from molecular dynamics simulations from a direct evaluation of the time integral of the heat-flux autocorrelation function  \cite{schelling2002comparison,carbogno2017ab,marcolongo2016microscopic,baroni2020heat}. However, we discussed here the derivation of the current-current correlation function in a certain diagrammatic approximation, and we further compared results obtained by different heat fluxes in the LSFA. The question then reduces to what is usually called a ``conserving approximation'' within the language of electrical transport, \textit{i.e}.\ an approximate result that satisfies the gauge-invariant requirements of the theory. So far, it is not evident that also for thermal transport such an approach can be established. Nonetheless, we cannot help noting that the Hardy choice for the heat flux includes an antiresonant term for the heat-flux operator that is quantitatively irrelevant for the thermal conductivity. As a consequence, the Wigner current has certainly the advantage to include from the beginning only the relevant interband processes for thermal transport, avoiding the computational effort to include irrelevant terms.

\section{Conclusions}
\label{sec:conclusion}
In summary, in this work, we have shown how to derive an expression for the thermal conductivity with a full quantum-mechanical many-body formalism based on Green's functions and the Kubo formula. We have focused on two definitions of the {energy field}, that lead via the continuity equation to two different definitions of the heat flux. These two choices have been motivated by previous work in the literature. The first one corresponds to the heat flux originally proposed by Hardy  \cite{hardy1963energy}, which follows from a coarse-graining procedure implemented on the local harmonic Hamiltonian in real space. 
The second one corresponds to the Wigner heat flux implemented within the approach of Refs.\  \cite{simoncelli2019unified,long_paper}. 
Once the heat flux is defined in terms of creation and annihilation phonon operators, one can derive the thermal conductivity as the zero-frequency limit of a current-current response function, in close analogy with the usual diagrammatic approach implemented for electric transport  \cite{mahan2013many}. Here, we compute the response function in the so-called dressed-bubble approximation, which neglects vertex corrections but accounts for all self-energy corrections of the phonon Greens' function due to interactions. Such an approximation allows us to derive an analytical expression for the thermal conductivity in terms of the exact phonon spectral density,  that can be computed in principle at any perturbative order and extended to scattering sources of any nature. Thus, the present result based on the full phonon spectral density can also be used in the overdamped regime of thermal transport, where the quasiparticle picture of phonons breaks down. 
Neglecting vertex corrections in the dressed-bubble approximation is the main limitation of the present work, that in principle cannot be applied in the hydrodynamic regime of thermal transport, in which heat is mainly carried by collective excitation of phonon \cite{cepellotti2016thermal}. Modeling hydrodynamic thermal transport requires {to account for} repopulation terms {in the description of scattering}  \cite{fugallo2014thermal,cepellotti2015phonon,cepellotti2016thermal,physrevx.10.011019}, {and is a possible future development of the present work}.

The thermal conductivity is given, for both choices of the heat flux, by an intraband and an interband contribution. The former is independent of the choice of the heat flux, while the second displays some differences both in the definition of interband velocities and in the way interband processes are weighted, with the appearance in Hardy's formulation of antiresonant terms. 
The present expression of the thermal conductivity in terms of the full phonon spectral densities derived from Hardy's heat flux coincides with the one presented in the theoretical studies  \cite{maradudin1964lattice,semwal1972thermal}. The analogous expression derived with Wigner heat flux is instead {a novel and} original result.  

With the LSFA applied to the full quantum formula \eqref{k wbte} {with the Wigner heat flux} we have {recovered} the results {of the} Wigner transport equation {derived} in Refs.\  \cite{simoncelli2019unified,long_paper}, thus proving the consistency of the Wigner formalism \textcolor{black}{ also in a fully quantum approach.} On the other hand, in the case of the Hardy heat flux we {have} found formal differences with respect to the quantum derivation recently proposed in Ref.\  \cite{isaeva2019modeling}. {We have argued that these differences originate from the procedure employed in Ref.\  \cite{isaeva2019modeling,pereverzev2018theoretical} to implement the limit of frequency-independent phonon lifetimes within the real-time Green's function,
leading to some differences in the interband terms (the intraband are instead identical) that yield, however, no quantitative differences when computed numerically in the materials tested so far. }

As benchmark examples, we have studied two systems with ultralow thermal conductivity, namely the perovskite \ch{CsPbBr3} and the zirconate \ch{La2Zr2O7}. As discussed, we did not find so far a satisfactory general argument to understand the quantitative agreement between different formulations of the heat flux. On the other hand, we argued that the Wigner formulation appears the most natural choice to compute the thermal conductivity in the dressed-bubble approximation, which reduces to the result from the Wigner transport equation in the LSFA. Indeed, it only includes the interband contributions which are actually relevant in this regime, at least for the systems tested so far. In the aforementioned test case materials, we have considered the effect of anharmonicity in the form of finite phonon lifetime due to 3-phonon and natural-abundance isotope scattering at the lowest perturbative order. Some of us in Ref.\ \cite{long_paper} have presented a comparison between Raman scattering data for \ch{CsPbBr3} and \ch{La2Zr2O7} and simulated Raman spectra within the same approximation, \textit{i.e}.\ considering lowest-order 3-phonon anharmonicity and no frequency renormalization, finding remarkable agreement between the two in the temperature range $100{-}300\,$K. This gives us reason to believe that  this level of approximation for the anharmonicity is accurate enough for the scope of the present study. In general, a comparison between the phonon frequencies and their linewidth obtained from ab-initio calculations and the phonon spectra obtained from inelastic neutron scattering \cite{lanigan2021two} or electron energy loss spectroscopy \cite{senga2019position}, could provide a quantitative evaluation  of the accuracy of the approximation done in considering phonon anharmonicity at the lowest non-trivial order for the whole phonon dispersion (Raman scattering is limited to $\bm q {=}0$). As a possible future perspective, it would be interesting to consider the effect of higher-order anharmonicity such as 4-phonon scattering on the results for thermal conductivity for these materials.

It may be interesting to search for a test case in which the details of different formulations of thermal conductivity might provide appreciable quantitative differences. Indeed, the identification of such discrepancies could help to understand the differences between the various formulations. From a broader perspective, it would be interesting to test the general quantum formula of thermal conductivity in terms of the full phonon spectral densities \eqref{k hardy}-\eqref{k wbte} derived in this paper. In the overdamped regime of lattice heat conduction, strong interactions may result in very broad (or non-Lorentzian) profiles for the spectral densities  \cite{aseginolaza2019phonon,lanigan2021two}. This could lead to significant mixing and coherence effects between many phonon bands far apart in energy. In this case, the difference between the definition of the off-diagonal velocity \eqref{hardy 2 wigner} --- that constitute most of the difference between the results from Hardy and Wigner heat fluxes --- should be enhanced, and the different formulations of interband thermal conductivity must reflect that in a clear discrepancy.
{The recent work of Ref.\  \cite{dangic2020origin} goes in this direction, even if the results for thermal conductivity obtained implementing the expression derived in  Ref.\  \cite{semwal1972thermal} for the test case \ch{GeTe} differ less than $15\% $ from the result of thermal conductivity computed with BTE in the SMA, whereas in complex crystals such discrepancy is typically far more remarkable, consisting up to $70\%$ of total conductivity for the materials we tested.  } 
All these issues provide an interesting perspective for future work on lattice thermal transport.

\begin{acknowledgements}
	This work has been supported by PRIN 2017 No.
	2017Z8TS5B, by Sapienza University via Grant No.
	RM11916B56802AFE and RM120172A8CC7CC7
and by EU under project MORE-TEM ERC-SYN (grant agreement No 951215).
	N.M. acknowledges support from the Swiss National Science Foundation (SNSF) and the MARVEL NCCR.
	M.S. acknowledges support from  SNSF project P500PT\_203178.
\end{acknowledgements}

\bigskip
\appendix
\onecolumngrid
\section{Details on the derivation of heat-flux operator}
\label{app:heat flux}
In this Appendix we show how to derive the expression for the heat flux associated with a given local energy operator, following Hardy's original procedure  \cite{hardy1963energy}. Its implementation for the Hardy local operator \eqref{hardy h(R)} has been detailed in Ref.\  \cite{hardy1963energy}. Here we will {employ this procedure to derive the heat flux starting from the local energy operator emerging from the Wigner formalism.}  This will also serve us to show how the novel definition of the velocity matrix used in Wigner formalism follows from the derivation of the heat flux. Let us start {by recalling} the Wigner local energy given in Eq.\ \eqref{e(R) WBTE}
\begin{equation} \label{e(R) WBTE app}
	\hat{\tenscomp{h}}(\bm R)_{b\a} {=} \frac{\h}{2}\!\!\sum_{\bm R\p\! b\p\!\a\p}\!\!\!\!\sqrt{\tenscomp{G}}_{\bm R b\a,\bm R\p b\p\!\a\p}\hat{\tenscomp{a}}^\dagger(\bm R)_{b\a}\hat{\tenscomp{a}}(\bm R\p)_{b\p\!\a\p} {+}\text{h.c.} ,
\end{equation}
which satisfies $\sum_{\bm{R} b \a}\hat{\tenscomp{h}}(\bm{R})_{b\a}{=}\hat{\mathcal H}$ (without the zero-point-motion term, which {does not affects the dynamics of the systems and thus does not} provide any heat flux). To derive the heat flux from the continuity equation \eqref{continuity}, one needs to define an energy density of continuous variable $h(\bm x)$. This can be done convolving the discrete operator \eqref{e(R) WBTE app} with a continuous normalized distribution $\Delta_\ell(\bm{x}{-}\bm{R}{-}\bm{\bm{\tau}}_b)$, such that
$\hat{h}(\bm{x}){=}\sum_{\bm{R}b\a}\hat{\tenscomp{h}}(\bm{R})_{b\a}\Delta_\ell(\bm{x}{-}\bm{R}{-}\bm{\bm{\tau}}_b)$.
It is easy to show that the continuous energy field satisfies $\int_V\dd[3]{x}\hat{h}(\bm{x}){=}\hat{\mathcal H}$. From the continuity equation \eqref{continuity}, one has
\begin{equation}
	{-}\nabla{\cdot}\hat{\bm{j}}(\bm{x})=\frac{d\hat{h}(\bm{x})}{dt}=-\frac{i}{\hbar}\big[\hat{h}(\bm{x}),\hat{\mathcal H}\big],
\end{equation}
implying that the divergence of the energy flux operator can be computed from the commutator
\begin{equation}
	\begin{split} 
		\nabla{\cdot}\hat{\bm{j}}(\bm{x}){=}
		\frac{i}{\hbar}\!\!\!\sum_{\substack{\bm{R}b\a \\ \bm{R}\p\!b\p\! \a\p} }\!\!\!\!\Big[\Delta_\ell(\bm{x}{-}\bm{R}{-}\bm{\bm{\tau}}_b ){-}\Delta_\ell(\bm{x}{-}\bm{R'}{-}\bm{\bm{\tau}}_{b\p})\big]\hat{\tenscomp{h}}(\bm{R})_{b\a}\hat{\tenscomp{h}}(\bm{R}\p)_{b\p\!\a\p}.
	\end{split}
	\raisetag{3mm}
	\label{eq:to_rewrite}
\end{equation}
Now the quantity in square bracket can be written in terms of the Taylor series of the distributions $\Delta_\ell$ around the point $\bm{x}$:
	\begin{equation}
		\begin{split}
			\Delta_\ell(\bm{x}{-}\bm{R}{-}\bm{\bm{\tau}}_{b}){-}\Delta_\ell(\bm{x}{-}\bm{R'}{-}\bm{\bm{\tau}}_{b\p})&=
			\sum_{|\gamma|}\tfrac{1}{\gamma!}
			\Big[\tfrac{\partial^\gamma }{\partial \bm{y}^\gamma} \Delta_\ell(\bm{y})\big|_{\bm{y}=\bm{x}} \Big](\bm{x}{-}\bm{R}{-}\bm{\bm{\tau}}_{b})^\gamma
			{-}
			\sum_{|\gamma|}\tfrac{1}{\gamma!}
			\Big[\tfrac{\partial^\gamma }{\partial \bm{y}^\gamma} \Delta_\ell(\bm{y})\big|_{\bm{y}=\bm{x}} \Big](\bm{x}{-}\bm{R'}{-}\bm{\bm{\tau}}_{b\p})^\gamma\\
			& 
			=\sum_{|\gamma|{\geq}1}\tfrac{1}{\gamma!}
			[(\bm{x}{-}\bm{R}{-}\bm{\bm{\tau}}_{b})^\gamma-(\bm{x}{-}\bm{R'}{-}\bm{\bm{\tau}}_{b\p})^\gamma]\tfrac{\partial^\gamma \Delta_\ell(\bm{x})}{\partial \bm{x}^\gamma}, 
		\end{split}
		\raisetag{9mm}
	\end{equation}
where $\gamma$ is a multiindex used for multi-dimensional Taylor expansions, and clearly the zeroth order term of the resulting series is zero.
It follows that Eq.\ (\ref{eq:to_rewrite}) can be recast as
	\begin{equation}
		\begin{split} 
			\nabla{\cdot}\hat{\bm{j}}(\bm{x}){=}
			\frac{i}{\hbar}\!\!\!\sum_{\substack{\bm{R}b\a \\ \bm{R}\p\!b\p\! \a\p} }\!\hat{\tenscomp{h}}(\bm{R})_{b\a}\hat{\tenscomp{h}}(\bm{R}\p)_{b\p\!\a\p} \!\!\!\!
			\sum_{|\gamma|{\geq}1}\!\!\tfrac{1}{\gamma!}
			[(\bm{x}{-}\bm{R}{-}\bm{\bm{\tau}}_{b})^\gamma{-}(\bm{x}{-}\bm{R'}{-}\bm{\bm{\tau}}_{b\p})^\gamma]\tfrac{\partial^\gamma \Delta_\ell(\bm{x})}{\partial \bm{x}^\gamma},
		\end{split}
		\raisetag{3mm}
		\label{eq:to_rewrite2}
	\end{equation}
	implying that the local heat-flux operator (\textit{i.e}.\ the current density) reads
	\begin{equation}
		\begin{split} 
			\hat{\bm{j}}(\bm{x}){=}&
			\frac{i}{\hbar}\!\!\!\sum_{\substack{\bm{R}b\a \\ \bm{R}\p\!b\p\! \a\p} }\!\hat{\tenscomp{h}}(\bm{R})_{b\a}\hat{\tenscomp{h}}(\bm{R}\p)_{b\p\!\a\p} 
			\Big\{
			(\bm{R'}{+}\bm{\bm{\tau}}_{b\p}{-}\bm{R}{-}\bm{\bm{\tau}}_{b}) \Delta_\ell(\bm{x}) 
			{+}\sum_{|\gamma|{\geq}2}\!
			\!\!\tfrac{1}{\gamma!}
			\!
			[(\bm{x}{-}\bm{R}{-}\bm{\bm{\tau}}_{b})^\gamma{-}(\bm{x}{-}\bm{R'}{-}\bm{\bm{\tau}}_{b\p})^\gamma]\tfrac{\partial^{\gamma-1} \Delta_\ell(\bm{x})}{\partial \bm{x}^{\gamma-1}}\Big\}.
		\end{split}
		\label{eq:to_rewrite2}
	\end{equation}
	For the computation of the thermal conductivity one needs the total heat flux, obtained integrating the density and normalizing by the {crystal's volume $V=\mathcal{V}{N_c}$ (where $\mathcal{V}$ is the volume of the primitive cell and $N_c$ the number of primitive cell in the crystal)}
	\begin{equation}
		\hat{\tens{J}}=\frac{1}{N_c\mathcal{V}}\int\limits_V \hat{\bm{j}}(\bm{x}) \dd[3]x.
	\end{equation}
	When integrating Eq.\ (\ref{eq:to_rewrite2}) over the volume, one has that $\Delta_\ell(\bm{x})$ is non-zero only over a small region of space, so the volume integral of the derivatives can be neglected and one obtains the expression for the average heat flux:
	\begin{equation}\label{J=[h,H]}
		\begin{split} 
			\hat{\tens{J}}=&
			\frac{-i}{\hbar N_c\mathcal{V}}\!\!\sum_{\substack{\bm{R}b\a \\ \bm{R}\p\!b\p\! \a\p} }\!(\bm{R}{+}\bm{\bm{\tau}}_{b}{-}\bm{R'}{-}\bm{\bm{\tau}}_{b\p})\hat{\tenscomp{h}}(\bm{R})_{b\a}\hat{\tenscomp{h}}(\bm{R}\p)_{b\p\!\a\p} =\frac{-i}{\hbar N_c\mathcal{V}}\!\!\sum_{\bm{R}b\a  }\!(\bm{R}{+}\bm{\bm{\tau}}_{b})[\hat{\tenscomp{h}}(\bm{R})_{b\a},\hat{\mathcal{H}} ]
			\:.
		\end{split}
	\end{equation}
	At this point we can plug in the commutator the definition of the local energy \eqref{e(R) WBTE app}, obtaining
	\begin{equation}\label{commutator [h(R),H]}
		\begin{split}
			&[\hat{\tenscomp{h}}(\bm{R})_{b\a},\hat{\mathcal{H}} ]= \!\!\tfrac{\h^2}{2}\hspace{-0.6cm}\sum_{ \substack{  \bm R\p\! b\p\!\a\p\!,  \bm R_1 b_1\a_1, \\ \bm R_2b_2 \a_2} }\hspace{-0.8cm} \sqrt{\!\tenscomp{G}}_{\bm Rb\a,\bm R_1 b_1 \a_1}\!\!\sqrt{\!\tenscomp{G}}_{\bm R\p\! b\p\!\a\p\!,\bm R_2 b_2 \a_2}[ \hat{\tenscomp{a}}^\dagger(\bm R)_{b\a}\hat{\tenscomp{a}}(\bm R_1)_{b_1\a_1}, \hat{\tenscomp{a}}^\dagger(\bm R\p)_{b\p\!\a\p}\hat{\tenscomp{a}}(\bm R_2)_{b_2\a_2} ] + \text{h.c.}\\
			&= \!\!\tfrac{\h^2}{2}\hspace{-0.6cm}\sum_{ \substack{  \bm R\p\! b\p\!\a\p\!,  \bm R_1 b_1\a_1, \\ \bm R_2b_2 \a_2} }\hspace{-0.8cm}  \sqrt{\!\tenscomp{G}}_{\bm Rb\a,\bm R_1 b_1 \a_1}\!\!\sqrt{\!\tenscomp{G}}_{\bm R\p\! b\p\!\a\p\!,\bm R_2 b_2 \a_2}\Big\{\hat{\tenscomp{a}}^\dagger(\bm R)_{b\a}\hat{\tenscomp{a}}(\bm R_2)_{b_2\a_2} \delta_{\bm R\p\bm R_1}\delta_{b\p b_1}\delta_{\a\p\a_1}  {-}\hat{\tenscomp{a}}^\dagger(\bm R\p)_{b\p\!\a\p} \hat{\tenscomp{a}}(\bm R_1)_{b_1\a_1}\delta_{\bm R\bm R_2}\delta_{b b_2}\delta_{\a\a_2} \Big\}{+} \text{h.c.} \:.
		\end{split}
	\end{equation}
	Now we can plug the latter expression in the heat flux \eqref{J=[h,H]}, and performing a relabelling $\bm R_2{\to}\bm R_1$ in the first term and $\bm R {\leftrightarrow }\bm R\p$ in the second term of the last line of \eqref{commutator [h(R),H]} we get
	\begin{equation}
		\begin{split}
			\hat{\tens{J}}{=}\frac{-i\h}{2N_c\mathcal{V}}\hspace{-0.2cm}
			\sum_{ \substack{  \bm R b\a,   \bm R\p\! b\p\!\a\p\!,\\ \bm R_1b_1 \a_1} }
			\hspace{-0.4cm}(\bm{R}{+}\bm{\bm{\tau}}_{b}{-}\bm{R'}{-}\bm{\bm{\tau}}_{b\p})\sqrt{\!\tenscomp{G}}_{\bm Rb\a,\bm R\p b\p \a\p}\!\sqrt{\!\tenscomp{G}}_{\bm R\p\! b\p\!\a\p\!,\bm R_1 b_1 \a_1}  \hat{\tenscomp{a}}^\dagger(\bm R)_{b\a}\hat{\tenscomp{a}}(\bm R_1)_{b_1\a_1} + \text{h.c.} 
		\end{split}
	\end{equation}
	Now {we rewrite the bosonic operators in real space in terms of their Fourier transform  \cite{long_paper},}
	\begin{equation}
		\label{eq:FT_op_a}
		\hat{\tenscomp{a}}(\bm R)_{b\a}=\frac{1}{\sqrt{N_c}}\sum_{\bm{q}}\hat{\tenscomp{a}}(\bm q)_{b\a}e^{+i\bm{q}\cdot(\bm{R}+\bm{\tau}_b)},
	\end{equation}
	obtaining
	\begin{equation}
		\begin{split}
			\hat{\tens{J}}&{=}\tfrac{-i\h}{2N_c\mathcal{V}}
			\sum_{\bm q}\hspace{-0.2cm}\sum_{ \substack{  b\a,   b\p\!\a\p\!,\\ b_1 \a_1} }\hspace{-0.2cm} \hat{\tenscomp{a}}^\dagger(\bm q)_{b\a}\tfrac{1}{\sqrt{N_c}}\!\sum_{\bm R\p\!\bm R_1}\hat{\tenscomp{a}}(\bm R_1)_{b_1\a_1}\sqrt{\!\tenscomp{G}}_{\bm R\p\! b\p\!\a\p\!,\bm R_1 b_1 \a_1} \!\sum_{\bm R}(\bm{R}{+}\bm{\bm{\tau}}_{b}{-}\bm{R'}{-}\bm{\bm{\tau}}_{b\p}) \sqrt{\!\tenscomp{G}}_{\bm Rb\a,\bm R\p b\p \a\p}\!e^{-i\bm q \vdot(\bm{R}{+}\bm{\bm{\tau}}_{b})}  + \text{h.c.} \\
			&{=}\tfrac{-i\h}{2N_c\mathcal{V}}\!\!\sum_{\bm q}\hspace{-0.2cm}\sum_{ \substack{  b\a,   b\p\!\a\p\!,\\ b_1 \a_1} }\hspace{-0.2cm} \hat{\tenscomp{a}}^\dagger(\bm q)_{b\a}\tfrac{1}{\sqrt{N_c}}\!\sum_{\bm R\p\!\bm R_1}\hat{\tenscomp{a}}(\bm R_1)_{b_1\a_1}\sqrt{\!\tenscomp{G}}_{\bm R\p\! b\p\!\a\p\!,\bm R_1 b_1 \a_1} e^{-i\bm q \vdot(\bm{R}\p{+}\bm{\bm{\tau}}_{b\p})}\!\sum_{\bm L}(\bm{L}{+}\bm{\bm{\tau}}_{b}{-}\bm{\bm{\tau}}_{b\p}) \sqrt{\!\tenscomp{G}}_{\bm L b\a,0 \p b\p \a\p}\!e^{-i\bm q \vdot(\bm{L}{+}\bm{\bm{\tau}}_{b}-\bm{\tau}_{b\p})}\! {+} \text{h.c.} \\
			&{=}\frac{\h}{2N_c\mathcal{V}}\sum_{\bm q}\hspace{-0.1cm}\sum_{ \substack{  b\a,   b\p\!\a\p\!,\\ b_1 \a_1} }\hspace{-0.2cm} \grad_{\!\!\bm q}\!\sqrt{\!\tenscomp{D}(\bm q)}_{b\a,b\p\!\a\p}\hat{\tenscomp{a}}^\dagger(\bm q)_{b\a}\frac{1}{\sqrt{N_c}}\!\sum_{\bm R\p\!\bm R_1}\hat{\tenscomp{a}}(\bm R_1)_{b_1\a_1}\sqrt{\!\tenscomp{G}}_{\bm R\p\! b\p\!\a\p\!,\bm R_1 b_1 \a_1} e^{-i\bm q \vdot(\bm{R}\p{+}\bm{\bm{\tau}}_{b\p})}  + \text{h.c.} 
		\end{split}
	\end{equation}
where we have multiplied and divided by $e^{-i\bm q \vdot(\bm{R}\p{+}\bm{\bm{\tau}}_{b\p})}$ and used the symmetry property \eqref{prop phi} of the matrix $\sqrt{\!\tenscomp{G}}_{\bm R b\a,\bm R\p\! b\p\! \a\p\!}{=}\sqrt{\!\tenscomp{G}}_{\bm R{-}\bm R\p\!b \a,0 b\p\! \a\p}$ considering a change of summation index  $\bm R {\to}\bm L{=}\bm R {-}\bm R\p$ to introduce the gradient of the dynamical matrix from \eqref{dynmat}. In the same fashion we can {recast} $\hat{\tenscomp{a}}(\bm R_1)$ in {terms of its reciprocal-space representation~(\ref{eq:FT_op_a})} to solve the summations in $\bm R\p, \bm R_1$ obtaining
\begin{equation}\label{J = a(q)b a(q)b}
	\begin{split}
		\hat{\tens{J}}{=}\frac{\h}{2N_c\mathcal{V}}\!\sum_{\bm q}\hspace{-0.2cm}\sum_{ \substack{  b\a,   b\p\!\a\p\!,\\ b_1 \a_1} }\hspace{-0.2cm} \grad_{\!\!\bm q}\!\sqrt{\!\tenscomp{D}(\bm q)}_{b\a,b\p\!\a\p}\!\!\sqrt{\!\tenscomp{D}(\bm q)}_{b\p\!\a\p\!,b_1 \a_1}\hat{\tenscomp{a}}^\dagger(\bm q)_{b\a}\hat{\tenscomp{a}}(\bm q)_{b_1\a_1}+ \text{h.c.}   
	\end{split}
\end{equation}
At this point we can pass to the normal mode basis {(Eq.\ (\ref{eq:ph_annihilation}))} using the transformation $\hat{\tenscomp{a}}(\bm q)_{b\a}{=}\sum_s\mathcal{E}(\bm q)_{s,b\a}\hat a(\bm q)_s$ while considering the definition and properties of the velocity matrix \eqref{velocity WBTE}, obtaining
\begin{equation}\label{J WBTE app}
	\begin{split}
		\hat{\tens{J}}{=}\frac{\h}{2N_c\mathcal{V}}\!\sum_{\bm q,ss\p}&\{ \w(\bm q)_s \tens{v}(\bm q)_{ss^\prime} \hat a^\dagger(\bm q)_s \hat a(\bm q)_{s^\prime}
		{+}\w(\bm q)_s \tens{v}^*(\bm q)_{ss^\prime} \hat a^\dagger(\bm q)_{s\p} \hat a(\bm q)_{s}\}\\
		=\frac{\h}{N_c\mathcal{V}}\!\sum_{\bm q,ss\p}&\frac{\omega(\bm q)_{s}\! +\! \omega(\bm q)_{s\p}}{2}   \tens{v}(\bm q)_{ss^\prime} \hat a^\dagger(\bm q)_s \hat a(\bm q)_{s^\prime}
	\end{split}
\end{equation}
which is the heat-flux operator introduced in  Eq.\ \eqref{J WBTE}. As mentioned above, the procedure for the derivation of the heat flux from Hardy energy density \eqref{hardy h(R)} is completely analogous and it is reported in Hardy's original paper  \cite{hardy1963energy}.

\section{{Gauge-invariance principle for the thermal conductivity: quantum considerations and implications on Hardy's derivation of the heat flux} } \label{app:gauge}
In this Appendix {we report some considerations on the application of the gauge-invariance principle for the thermal conductivity  \cite{ercole2016gauge} in a quantum-mechanical framework, and we show how these allow to simplify the expression for the heat flux operator obtained following Hardy's approach. }
{The starting point of the aforementioned work is the observation that energy's extensive character} implies the energy density to be defined up to a divergence of a bounded vector field, \textit{i.e}.\ $\hat h\p(\bm x,t){=}\hat h(\bm x,t) {+} \grad {\vdot} \hat{\bm p}(\bm x,t) $. Indeed, if $\hat{\bm{p}}(\bm x,t)$ is bounded, the volume integral of $\hat h\p(\bm x,t)$ --- \textit{i.e}.\ the energy --- differs from the volume integral of $\hat h(\bm x)$ by a term that scales as a surface, hence is subextensive and does not contribute to the energy in the thermodynamical limit  \cite{marcolongo2016microscopic}. As a consequence, it follows an indeterminacy on the current density, since the two are related by the continuity equation \eqref{continuity}. The indeterminacy on energy and current density thus acquire the structure of the gauge indeterminacy of vector and scalar potential of the electromagnetic field, namely
\begin{equation}\label{gauge indeterminacy}
	\begin{split}
		&\begin{cases}
			\hat h\p(\bm x,t)=\hat h(\bm x,t) +  \grad {\vdot} \hat{\bm p}(\bm x,t) \\
			\hat{\bm{j}}\p(\bm x,t)  = \hat{\bm{j}}(\bm x,t) - \dv{}{t}\hat{\bm p} (\bm x,t)
		\end{cases}\\
		&\hat{\bm{J}}\p(t)  = \hat{\bm{J}}(t) - \dv{}{t}\hat{\bm P} (t)
	\end{split}
\end{equation}
in which the last line describes the gauge transformation for the macroscopic heat flux, and $\hat{\bm P} {=}\tfrac{1}{N_c\mathcal{V}}\int_V\! \dd[3]{\!x}\hat{\bm p}(\bm x) $. The gauge invariance principle states that the indeterminacy \eqref{gauge indeterminacy} on the heat flux does not affect the thermal conductivity, so that
\begin{equation}
	\kappa\p{=} \tfrac{\mathcal{V}}{ k_BT^2} \int\limits_0^\infty\!\!\dd{t} \langle\hat{\bm {J}}\p(t)\!\vdot\!\hat{\bm{J}}\p(0)\rangle {=}\tfrac{\mathcal{V}}{ k_BT^2} \int\limits_0^\infty\!\!\dd{t} \langle\hat{\bm {J}}(t)\!\vdot\!\hat{\bm{J}}(0)\rangle {=} \kappa,
\end{equation}
hence the thermal conductivity is well-defined regardless of the indeterminacy \eqref{gauge indeterminacy} on the microscopic energy and current density. The gauge invariance of thermal conductivity was proven originally in a classical statistical mechanics framework; in this Appendix we will extend such proof to the quantum mechanical picture and we will discuss its capabilities in the case of the Hardy heat-flux operator. 
\subsection{{Quantum-mechanical considerations on the gauge invariance of thermal conductivity}}
{To show the} gauge invariance of thermal conductivity in quantum-mechanical picture, let us rephrase the Kubo formula \eqref{kubo lambda}
using the definition of the current-current correlation function introduced in Eq.\ \eqref{S = JJ}.
{Inserting} \eqref{S = JJ} in \eqref{kubo lambda} without resorting to {the} fluctuation-dissipation theorem, we get
\begin{equation}
	\kappa^{\alpha\beta} = \tfrac{N_c\mathcal{V}}{ k_BT^2}S^{\a\b}(\w {=} 0) {=} \tfrac{N_c\mathcal{V}}{k_BT^2}\!\int\limits_{-\infty}^{+\infty}\!\!\! \dd{t} \langle\hat{J}^{\a}(t)\hat{J}^{\b}(0)\rangle \:.    
\end{equation}
In order to {have} gauge invariance, {we have to verify that the following condition is satisfied:}
\begin{equation}\label{to prove gauge}
	\kappa\p\,\!^{\alpha\beta} {=} \tfrac{N_c\mathcal{V}}{k_BT^2}\!\int\limits_{-\infty}^{+\infty}\!\!\! \dd{t} \langle [\hat{J}^{\a}(t){-} \dv{\hat P^\a(t) }{t} ][\hat{J}^{\b}(0) {-}\dv{\hat{P}^\b(0)}{t} ]\rangle {=} \kappa^{\a\b} .    
\end{equation}
To this aim, we can show that any average that contains at least one total time derivative of an operator vanishes in \eqref{to prove gauge}. In fact, if we expand in the basis $\{\ket{n}\}$ of the eigenstates of the (full) Hamiltonian $\hat{H}$ defined such that $\hat{H}\ket{n}{=}E_n \ket{n}$, we get 
\begin{equation}
	\begin{split}
		\langle \hat{J}^{\a}(t) \dv{\hat{P}^\b(0)}{t}(0) \rangle {=}\sum\limits_{nm} \frac{e^{-\b E_n }}{\mathcal{Z}}\mel{n}{  \hat{J}^{\a}(t) }{m}\!\mel{m}{ \dv{\hat{P}^\b(0)}{t}(0) }{n},
	\end{split}
\end{equation}
where $\mathcal{Z} {=}\Tr[e^{-\b \hat{H}} ]$ is the canonical partition function. In the latter expression we can now use the Heisenberg scheme for the evolution of operators $i\h \dv{t} \hat{A}(t) {=} [\hat{H},\hat{A}(t)] $, which yields
\begin{equation} \label{J dP/dt spectral}
	\begin{split}
		\langle \hat{J}^{\a}(t) \dv{\hat{P}^\b(0)}{t}(0) \rangle {=}\frac{i}{\h}\sum\limits_{nm} \frac{e^{-\b E_n }}{\mathcal{Z}}e^{{-}\frac{i}{\h}(E_n{-}E_m)t } (E_n{-}E_m)\\ \times\mel{n}{\hat{J}^{\a}}{m}\!\mel{m}{ \hat{P}^\b }{n}.
	\end{split}
\end{equation}
Evidently, this term vanishes in the time integral of \eqref{to prove gauge}, since 
\[
\int\limits_{-\infty}^{+\infty}\!\!\! \dd{t}\!e^{{-}\frac{i}{\h}(E_n{-}E_m)t } \tfrac{(E_n{-}E_m)}{\h}{=}2\pi(E_n{-}E_m) \delta(E_n{-}E_m) {=}0
\]
and this proves \eqref{to prove gauge} and gauge invariance of thermal conductivity. Note that this result is valid with the assumption that the operator $\hat P $ is bounded, which guarantees that every matrix element of $\hat P $ in \eqref{J dP/dt spectral} is finite. \\
This is one of the key assumptions for the result of gauge invariance \cite{ercole2016gauge}; in fact, notice from Eq.\ \eqref{J=[h,H]} how the heat flux operator itself can be expressed as a total time derivative, namely the derivative of the first moment of the energy density $\hat{\bm{J}} {=}\dv{t} \sum\limits_{\bm R b \alpha} (\bm R + \bm \tau_b) \hat h(\bm R)_{b\alpha}$. However, the latter expression features a time derivative of an operator which is clearly unbounded in the thermodynamical limit, and hence can yield a finite conductivity.

\subsection{{Gauge invariance and simplification of Hardy's derivation of the heat flux}}
{Here we show how the gauge invariance properties of the thermal conductivity} can be used to {simplify significantly} the structure of Hardy's heat-flux operator. Let us start by recalling the local energy operator used by Hardy  \cite{hardy1963energy}, introduced in Eq.\ \eqref{hardy h(R)}
\begin{equation}
	\label{hardy h(R) app}
	\hat{h}(\bm R)_{b\a} = \frac{\hat{p}^2(\bm R)_{b\a}}{2M_{b}} + \hat V(\bm R)_{b\a}
\end{equation}
where we have expressed for brevity the harmonic ``on-site potential'' as $\hat V(\bm R)_{b\a}{=}\tfrac{1}{2} \bm \Phi_{\bm Rb\a, \bm R\p b\p \a\p} \hat{u}(\bm R)_{b\a}\hat{u}(\bm R\p)_{b\p\!\a\p} $. As discussed in Sec.~\ref{sec:heat flux}, the procedure to derive the heat-flux operator requires the convolution of the local-energy operator \eqref{hardy h(R)} with a smooth normalized distribution  $\Delta_\ell$ in order to define an energy density {depending on a} continuous variable, $\hat h(\bm x)$. We anticipated in the main text {that} there is an ambiguity on where to center the coarse-graining functions $\Delta_\ell$. Hardy originally considered the center of the $\Delta_\ell$ on the instantaneous position of the nuclei  $\hat{\bm r}(\bm R)_{b} = \bm R{+}{\bm \tau}_{b} {+}\hat{\bm{u}}(\bm R)_{b}$ \textit{i.e}.\ (the time dependence of the operators is omitted for brevity)
\begin{equation}
	\label{hardy h(x) original app}
	\hat{h}_{\text{Ref.\  \cite{hardy1963energy}}}(\bm x) {=}\frac{1}{2}\! \sum\limits_{\bm R b\a}\hat h(\bm R)_{b\a}\Delta_\ell(\bm x{-}\bm R{-}{\bm \tau}_{b} {-}\hat{\bm u}(\bm R)_{b\a} ) + \text{h.c.}
\end{equation}
(h.c.\ stands for hermitian conjugate) whereas in this work we have centered such smoothening functions in the equilibrium position of the nuclei, as described in Eq.\ \eqref{hardy h(x)}
\begin{equation}
	\label{hardy h(x) original app}
	\hat{h}(\bm x) {=} \sum\limits_{\bm R b\a}\hat h(\bm R)_{b\a}\Delta_\ell(\bm x {-} \bm R {-} \bm{\bm{\tau}}_b ) .
\end{equation}
Using \eqref{hardy h(x) original app} in the procedure described in Appendix \ref{app:heat flux} for the derivation of the heat flux leads to
\begin{equation}
	\begin{split}
		&\hat{\bm{J}}_{\text{Ref.\  \cite{hardy1963energy}}}{=} \frac{1}{2N_c\mathcal{V}}\!\sum_{\bm R b\a} \biggl\{\tfrac{\hat{\textbf{p}}(\bm R)_{b} }{M_{b}}\big( \tfrac{\hat{{p}}(\bm R)_{b\a}^2 }{2M_{b}} {+} \hat V(\bm R)_{b\a} \big) {+}
		\!\!\sum_{\bm R\p\! b\p\!\a\p}\! (\hat{\bm r}(\bm R)_{b}{-}\hat{\bm r}(\bm R\p)_{b\p} )  \frac{1}{i\hbar} \big[ \tfrac{ \hat p(\bm R)_{b\a}^2}{2M_{b}},\hat V(\bm R\p)_{b\p\!\a\p}\big] \biggr\}   +\text{ h.c.},
	\end{split}
	\label{J hardy original app}
\end{equation}
which is the original structure of the heat flux proposed by Hardy in  \cite{hardy1963energy}. The latter has to be compared with the one we used  in Eq.\ \eqref{J = up}, which we recast as
\begin{equation}
	\label{J = up app}
	\hat{\bm{J}}= \frac{1}{N_c\mathcal{V}}\sum_{\bm R b\a}  (\textbf{R}{+}\bm{\bm{\tau}}_b {-}\textbf{R}\p{-}\bm{\bm{\tau}}_{b\p})  \frac{1}{i\hbar} \big[ \tfrac{ \hat p(\bm R)_{b\a}^2}{2M_{b}},\hat V(\bm R\p)_{b\p\!\a\p}\big] \:.
\end{equation}
Evidently, the structure of the heat-flux operator \eqref{J hardy original app} is far more complicated than \eqref{J = up app}, and notably it features anharmonic $\mathcal{O}(p^3)$ and $\mathcal{O}(u^3)$ generated by an harmonic energy density such as \eqref{hardy h(R) app}. One could argue that \eqref{J hardy original app} is a richer description of the heat flux in a lattice, but actually we can 
{exploit the aforementioned properties} of gauge invariance  to prove that the heat fluxes \eqref{J hardy original app} and \eqref{J = up app} are related by a gauge transformation of the form \eqref{gauge indeterminacy}, hence produce the same result for thermal conductivity. To this aim let us rephrase the convective part of the heat flux \eqref{J hardy original app} as
\begin{equation}\label{convective}
	\tfrac{1}{2N_c\mathcal{V}}\!\sum_{\bm R b\a}\!\! \tfrac{\hat{\textbf{p}}(\bm R)_{b} }{M_{b}}\big( \tfrac{\hat{{p}}(\bm R)_{b\a}^2 }{2M_{b}} {+} \hat V(\bm R)_{b\a} \big) {=} \tfrac{1}{2N_c\mathcal{V}}\!\sum_{\bm R b\a}\tfrac{d\hat{\bm{u}}(\bm R)_b }{dt} \hat{h}(\bm R)_{b\a}
\end{equation}
while the conductive part of \eqref{J hardy original app} can be recast as
\begin{equation}\label{conductive}
	\begin{split}
		\tfrac{1}{2N_c\mathcal{V}}\!\!\sum_{\substack{\bm R b\a \\ \bm R\p\! b\p\!\a\p}}(\hat{\bm r}(\bm R)_{b}{-}\hat{\bm r}(\bm R\p)_{b\p} )\frac{1}{i\hbar} \big[ \tfrac{ \hat p(\bm R)_{b\a}^2}{2M_{b}},\hat V(\bm R\p)_{b\p\!\a\p}\big]
		&{=}\frac{1}{2} \hat{\bm{J}} {+} \tfrac{1}{2N_c\mathcal{V}}\!\!\sum_{\substack{\bm R b\a \\ \bm R\p\! b\p\!\a\p}}(\hat{\bm u}(\bm R)_{b}{-}\hat{\bm u}(\bm R\p)_{b\p} )\frac{1}{i\hbar} \big[ \tfrac{ \hat p(\bm R)_{b\a}^2}{2M_{b}},\hat V(\bm R\p)_{b\p\!\a\p}\big]\\
		&{=}\frac{1}{2} \hat{\bm{J}} {+}\frac{1}{2N_c\mathcal{V}}\!\! \sum_{\bm R b\a}\hat{\bm u}(\bm R)_{b} \dv{\hat h(\bm R)_{b\a}}{t} \:.
	\end{split}
\end{equation}
Using \eqref{convective}-\eqref{conductive} in \eqref{J hardy original app} and solving the $\tfrac{1}{2}$ factors with the hermitian conjugate yields
\begin{equation}\label{J hardy = J + d/dt}
	\hat{\bm{J}}_{\text{Ref.\  \cite{hardy1963energy}}}{=}   \hat{\bm{J}} +\frac{1}{2N_c\mathcal{V}}\dv{t}[ \sum_{\bm R b\a}\hat{\bm u}(\bm R)_{b} \hat h(\bm R)_{b\a} ]
\end{equation}
which proves our point: centering the $\Delta_\ell$ smoothening functions on the instantaneous position of the ions as in \eqref{hardy h(x) original app} or in their equilibrium position as in \eqref{hardy h(R) app} leads to two heat-flux operators that differ by  total time derivative of a \textcolor{black}{ bounded} term. By virtue of the gauge invariance of thermal conductivity, the two heat fluxes  describe the same physics of heat transport and can be thought as two ``gauges'' of the same heat flux \cite{isaeva2019modeling}. Therefore, this invariance principle provides a solid theoretical reasoning to select the heat flux \eqref{J = up app} over \eqref{J hardy original app}, choice that greatly simplifies the calculations.\\
Note that the identification  $\tfrac{\hat{\bm p}(\bm R)_b}{M_b}{=} \derivative{\hat{\bm u}(\bm R)_b }{t}$ that  we used in \eqref{convective} implies the important assumption of no atomic diffusion, \textit{i.e}.\ no drifting of the ionic equilibrium position  $\dv{(\bm{R}{+}\bm{\bm{\tau}}_b ) }{t}{=}0$. This assumption, on which this whole proof holds, is typically met and causes no complications in a crystalline solid.\\

\section{Details on the derivation of thermal conductivity}
\label{app:conductivity}
In this Appendix we will describe how to derive {the} thermal conductivity {expression} discussed in \ref{sec:conductivity} using the formalism of finite-temperature Green's function. We will follow the standard many-body approach for computing DC conductivities, outlined e.g.\ in Mahan's textbook  \cite{mahan2013many}. Our goal is  to compute thermal conductivity from the retarded response function, using the relation \eqref{k = chi} derived from fluctuation-dissipation theorem 
\begin{equation}
	\label{k = chi app}
	\k^{\a\b}= \frac{N_c\mathcal{V}}{ T}\lim\limits_{\w \to 0} \frac{\Im \chi^{\a\b}(\w +i0^+)}{\w} \:.
\end{equation}
As anticipated, the starting point is the time-ordered current-current response function in the imaginary time $\tau$
\begin{equation}
	\label{chi(tau) app}
	\chi^{\a\b}(\tau) = \langle \mathcal{T}_{\tau} \hat{J}^{\a}(\tau)\hat{J}^{\b}(0) \rangle\:
\end{equation}
which we shall rewrite in terms of the explicit expression of the heat flux chosen. We recap the expressions of the heat-flux operators in Wigner and Hardy formalism
\begin{subequations}
	\begin{eqnarray}
		&&\hspace{2.5cm}\text{\textbf{Wigner} }\nonumber \\
		&& \hat{\tens{J}}{=}\frac{\h}{N_c\mathcal{V}}\!\sum_{\bm q,ss\p}\!\frac{\omega(\bm{q})_s{+}\omega(\bm{q})_{s'}}{2}{\tens{v}}(\bm{q})_{ss'}\hat{a}^\dagger\!({\bm{q}})_{s}
		\hat{a}(\bm{q})_{s'} \label{J wig app}\\
		&&\hspace{2.5cm}\text{\textbf{Hardy} }\nonumber \\
		&&\bm{\hat{J}} = \bm{\hat{J}}_R + \bm{\hat{J}}_A  \nonumber \\
		\label{J res app}
		&&\bm{\hat{J}}_R {=}\frac{\h}{N_c\mathcal{V}}\!\sum_{\bm q,ss\p}\!\frac{\omega(\bm q)_{s} \!+\! \omega(\bm q)_{s\p}}{2}  \bm v(\bm q)_{ss^\prime} \hat a^\dagger\!(\bm q)_s \hat a(\bm q)_{s^\prime}\quad \\
		\label{J ares app}
		&&\begin{split}
			\bm{\hat{J}}_A {=}\frac{\h}{N_c\mathcal{V}}\!\sum_{\bm q,ss\p}\!&\frac{\omega(\bm q)_{s}\! -\! \omega(\bm q)_{s\p}}{4} \bm v(\bm q)_{ss^\prime}\left[\hat a(-\bm q)_s \hat a(\bm q)_{s^\prime}\!-\!\hat a^\dagger(\bm q)_s\hat a^\dagger(-\bm q)_{s^\prime} \right]\!.
		\end{split}
	\end{eqnarray}
\end{subequations}
As we will show, the structure of the final result only depends on the combination of phonon creation/annihilation operators {appearing} in the {expression for} the heat flux. Therefore the calculation of the response function \eqref{chi(tau) app} for the resonant part of the Hardy heat flux will be equal to one for the Wigner heat flux, apart from the prefactor of the velocity matrices. With this reasoning, we will only perform the calculation for Hardy heat flux \eqref{J res app}-\eqref{J ares app}, and the result for Wigner heat flux will follow. Using the expressions of Hardy heat flux in \eqref{chi(tau) app} yields
	\begin{equation}\label{chi 2 part}   
		\begin{split}
			\chi^{\a\b}(\tau) =& \frac{\hbar^2}{N_c^2\mathcal{V}^2} \!\!\sum_{\bm q, ss\p}\!\sum_{\bm k, ll\p}\!\frac{\omega(\bm q)_{s} \!+\! \omega(\bm q)_{s\p}}{2}\frac{\omega(\bm q)_{l} \!+\! \omega(\bm q)_{l\p}}{2} v^\a(\bm q)_{ss^\prime} v^\b(\bm k)_{ll^\prime}\langle \mathcal{T}_{\tau} \hat a^\dagger\!(\bm q,\tau)_s \hat a(\bm q,\tau)_{s^\prime} \hat a^\dagger\!(\bm k)_l \hat a(\bm k)_{l^\prime}\rangle\\
			&+\frac{\hbar^2}{N_c^2\mathcal{V}^2} \!\!\sum_{\bm q, ss\p}\!\sum_{\bm k, ll\p}\!\frac{\omega(\bm q)_{s} \!-\! \omega(\bm q)_{s\p}}{4}\frac{\omega(\bm k)_{l} \!-\! \omega(\bm k)_{l\p}}{4} v^\a\!(\bm q)_{ss^\prime} v^\b\!(\bm k)_{ll^\prime}\\ &\hspace{3cm}\times \langle \mathcal{T}_{\tau}\left[\hat a({-}\bm q,\tau)_s \hat a(\bm q,\tau)_{s^\prime}{-}\hat a^\dagger(\bm q,\tau)_s\hat a^\dagger({-}\bm q,\tau)_{s^\prime} \right]\!\!\left[\hat a({-}\bm k)_l \hat a(\bm k)_{l^\prime}{-}\hat a^\dagger(\bm k)_l\hat a^\dagger(-\bm k)_{l^\prime} \right]  \rangle ,
		\end{split}
	\end{equation}
from which is evident how the response function is related to the two-particle phonon propagator  \cite{mahan2013many}. The two terms in \eqref{chi 2 part} represent respectively a resonant response function and an antiresonant response function, namely  $\chi_R^{\a\b}(\tau) = \langle \mathcal{T}_{\tau} \hat{J}^{\a}_R(\tau)\hat{J}^{\b}_R(0) \rangle$  and $\chi_A^{\a\b}(\tau) = \langle \mathcal{T}_{\tau} \hat{J}^{\a}_A(\tau)\hat{J}^{\b}_A(0) \rangle$. This separation is made possible by the fact that the resonant and antiresonant heat fluxes posses a different number of creation/annihilation phonon operators, hence mixed $\langle T_\tau \hat{J}_R(\tau)\hat{J}_A(0)\rangle$ terms are zero (at least at the present level of approximation). 
\begin{figure}
	\centering
	\includegraphics[keepaspectratio,width=0.7\columnwidth]{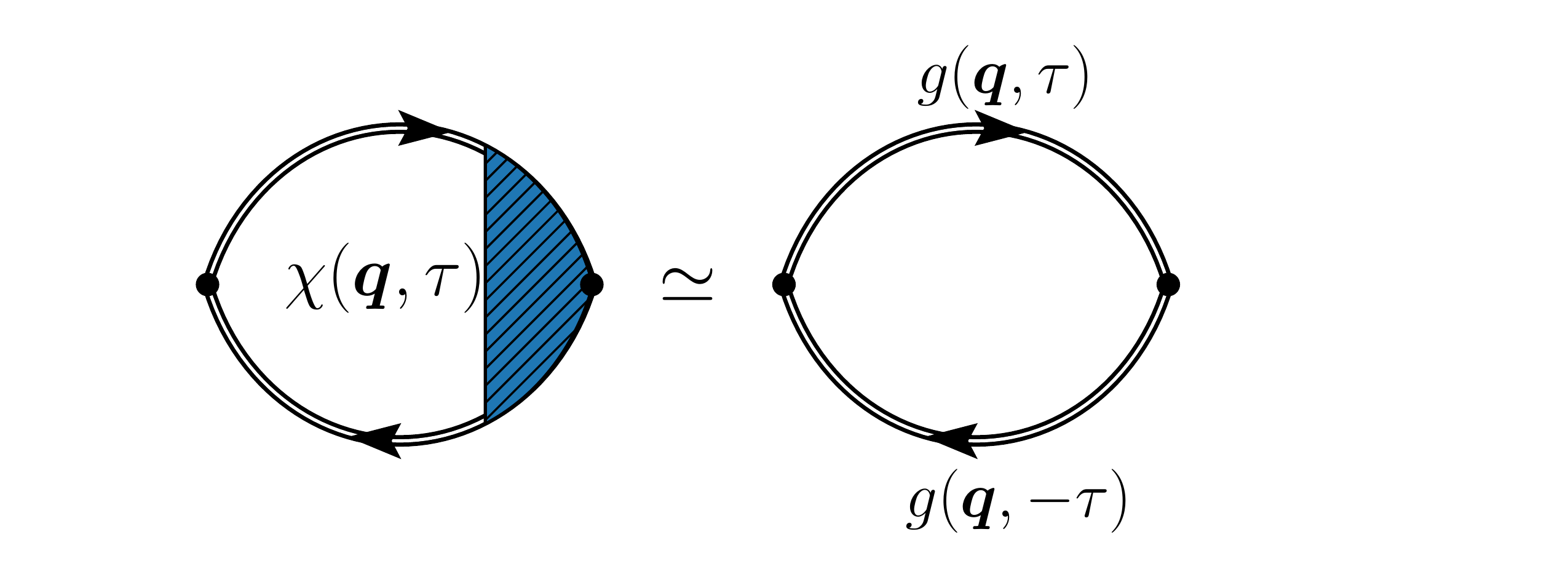}
	\caption{Diagramatic representation of dressed bubble approximation from Eq.\ \eqref{dressed bubble}. Left-hand side, full current-current correlation function $\chi(\bm q, \tau)$. In the right-hand side, double solid lines represent interacting phonon Green's functions $g$ from Eq.\ \eqref{g(t) app}.}
	\label{fig:dressed_bubble}
\end{figure}
More specifically, we will implement a dressed-bubble approximation, \textit{i.e}.\ we will consider only bubble diagrams where the phonon lines are fully dressed by interaction, as diagramatically represented in Fig.\ \ref{fig:dressed_bubble}. This amounts to neglecting vertex corrections and considering only self-energy effects in the phonon Green's functions, hence the two particle propagator is approximated as the product of two (interacting) phonon Green's functions, namely for the time-ordered product in the first term of \eqref{chi 2 part}
\begin{equation}
	\label{dressed bubble}
	\begin{split}
		\langle \mathcal{T}_{\tau}\hat a^\dagger\!(\bm q,\!\tau)_s \hat a(\bm q,\!\tau)_{s^\prime} \hat a^\dagger\!(\bm k)_l \hat a(\bm k)_{l^\prime}\rangle  {\simeq} g(\bm q,\!\tau)_{s\p} g(\bm q,\!\! -\tau)_{s} \! 
		\delta_{\bm q ,\bm k} \delta_{sl\p\!}\delta_{s\p\! l}, 
	\end{split}
\end{equation}
where the phonon Green's functions were introduced in \eqref{g(t)} as
\begin{equation}  \label{g(t) app}
	\begin{split}
		g(\bm q, \tau)_s\!& =\! -\langle \mathcal{T}_{\tau} \hat a(\bm q, \tau)_s \hat a^\dagger(\bm q, 0)_s  \rangle =
		\begin{cases}\! -\langle \hat a(\bm q, \tau)_s \hat a^\dagger\!(\bm q, 0)_s  \rangle  \quad \tau\!>\!0\\ \! -\langle  \hat a^\dagger\!(\bm q, 0)_s \hat  a(\bm q, \tau)_s\rangle  \quad \tau\!<\!0 .\end{cases}    
	\end{split}
\end{equation}
To be precise, we should also mention that the presence of $\delta_{ss\p}$ in \eqref{dressed bubble} consists in the additional approximation of neglecting branch mixed $\langle \hat a(\bm q)_s \hat a^\dagger(\bm q)_{s\p}\rangle$ lines, since these are higher order in the self-energy  \cite{maradudin1962scattering}. Using the approximation \eqref{dressed bubble} in \eqref{chi 2 part} evaluating all the possible pairwise contractions we get 
	\begin{equation} \label{chi res+ ares app}
		\begin{split}
			\chi^{\a\b}(\tau)& = \frac{\hbar^2}{N_c^2\mathcal{V}^2} \!\!\sum_{\bm q, ss\p}\!\frac{(\omega(\bm q)_{s} \!+\! \omega(\bm q)_{s\p})^2}{4}v^\a(\bm q)_{ss^\prime} v^\b(\bm q)_{s\p\! s} g(\bm q,\!\tau)_{s\p} g(\bm q,\!\! -\tau)_{s} \\
			&+\frac{\hbar^2}{N_c^2\mathcal{V}^2} \!\!\sum_{\bm q, ss\p}\!\frac{(\omega(\bm q)_{s} \!-\! \omega(\bm q)_{s\p})^2}{4} v^\a\!(\bm q)_{ss^\prime} v^\b\!(\bm q)_{s\p\! s}\tfrac{1}{2}\left[ g(\bm q,\!\tau)_{s\p} g(\bm q,\!\!\tau)_{s}{+}g(\bm q,\!-\tau)_{s\p} g(\bm q,\!\!-\tau)_{s} \right]    
		\end{split}
	\end{equation}
in which we have extensively used the property \eqref{velocity WBTE} of the velocity matrix  to symmetrize the result ({we stress that this property is verified by both Hardy's and Wigner's velocity matrices}). Now we can take the Fourier transform of \eqref{chi res+ ares app}, passing from imaginary-time to imaginary-frequency domain with $\chi^{\a\b}(i\Omega_n) {=}\hspace{-0.3cm} \int\limits_0^{{1}/{k_BT}}\hspace{-0.3cm}\dd \tau e^{i\Omega_n\tau} \chi^{\a\b}(\tau) $, which yields
	\begin{eqnarray} \label{chi res+ ares app}
		&& \chi^{\a\b}(i\Omega_n) = \frac{\h^2k_BT}{N_c^2\mathcal{V}^2}\! \!\!\sum_{\bm q, ss\p}\!\!\frac{(\omega(\bm q)_{s} \!+\! \omega(\bm q)_{s\p})^2}{4}v^\a(\bm q)_{ss^\prime} v^\b(\bm q)_{s\p\! s}\!\!\sum\limits_{\w_m} g(\bm q,i\Omega_n{+}i\w_m)_{s\p} g(\bm q, i\w_m)_{s} \nonumber\\
		&&+\frac{\h^2k_BT}{N_c^2\mathcal{V}^2}\! \!\!\sum_{\bm q, ss\p}\!\!\frac{(\omega(\bm q)_{s} {-} \omega(\bm q)_{s\p})^2}{4} v^\a\!(\bm q)_{ss^\prime} v^\b\!(\bm q)_{s\p\! s}\!\!\sum\limits_{\w_m} \tfrac{1}{2}\left[ g(\bm q,i\Omega_n{-}i\w_m)_{s\p} g(\bm q,i\w_m)_{s}{+}g(\bm q,{-}i\Omega_n{-}i\w_m)_{s\p} g(\bm q,i\w_m)_{s} \right], \hspace{1cm}
	\end{eqnarray}
where the Matsubara bosonic frequencies are defined as $i\Omega_n =  2\pi k_BT n,\: n\in \mathbb{Z}$. The next step {consists in expressing} the Green's function in terms of the spectral densities, using a Lehmann representation (see e.g.\ Mahan's textbook Ref.\  \cite{mahan2013many} sec. 3.3)
\begin{equation}
	\label{lehman app}
	g(\bm q,i\w_l)_s = \int\limits_{-\infty}^{+\infty}\dd{\w} \frac{b(\bm q,\w)_s}{i\w_l -\hbar\w} \:.
\end{equation}

Using  \eqref{lehman app} in \eqref{chi res+ ares app} yields an expression of the response function in the imaginary frequency domain in which the Matsubara frequency terms can be computed straightforwardly using standard summation techniques  \cite{mahan2013many}, that will yield a combination of Bose-Einstein occupation functions.
After having performed the Mastubara summation, we can consider the analytic continuation $i\Omega_n{\to} \hbar\w {+} i0^+$  obtaining the finite frequency response function, whose imaginary part reads
	\begin{equation}
		\begin{split}
			{\Im}\chi^{\a\b}(\w {+}i0^+) {=} \tfrac{\h^2\pi}{N_c^2\mathcal{V}^2} \!\!\sum_{\bm q, ss\p}\!\!
			\tfrac{(\omega(\bm q)_{s} {+} \omega(\bm q)_{s\p})^2}{4}
			v^{\a}(\bm q)_{ss^\prime} v^{\b}(\bm q)_{s\p\! s}
			\!\!\int\!\!\!\!\!\int\!\!\dd{\w_1}\!\dd{\w_2}b(\bm q,\w_1)_{s\p}b(\bm q,\w_2)_{s} \delta(\w_1{-}\w_2{-}\w) \tfrac{1}{\h} \left[{-}n_T(\w_1){+}n_T(\w_2) \right] \\
			{+} \tfrac{\h^2\pi}{N_c^2\mathcal{V}^2} \!\!\sum_{\bm{q}, ss\p}\!\!
			\tfrac{(\omega(\bm q)_{s} {-} \omega(\bm q)_{s\p})^2}{4} v^\a\!(\bm q)_{ss^\prime} v^\b\!(\bm q)_{s\p\!s} 
			\!\!\int\!\!\!\!\!\int\!\!\dd{\w_1}\!\dd{\w_2}b(\bm q,\w_1)_{s\p}b(\bm q,\w_2)_{s} 
			\tfrac{1}{2\hbar} \bigl\{ \delta(\w_1{+}\w_2{+}\w)\bigl[n_T({-}\w_1){-}n_T(\w_2) \bigr]\\
			{-} \delta(\w_1{+}\w_2{-}\w) \bigl[ n_T({-}\w_2){-}n_T(\w_1) \bigr] \bigr\}  \:.
		\end{split}
	\end{equation}
	Finally from the latter expression, we can easily perform the static limit \eqref{k = chi app} to obtain thermal conductivity, which reads 
	\begin{equation}\label{k hardy app}
		\begin{split}
			\kappa^{\a\b} = &\frac{\h^2\pi}{N_c\mathcal{V}T}\!\!\sum_{\bm q, ss\p}\!\!\frac{(\omega(\bm q)_{s} \!+\! \omega(\bm q)_{s\p})^2}{4}v^\a(\bm q)_{ss^\prime} v^\b(\bm q)_{s\p\! s}\!\!\int\!\dd{\w}b(\bm q,\w)_{s\p}b(\bm q,\w)_{s} \left[ {-}\pdv{n_T(\w)}{\hbar\w} \right] \\
			{+}&\frac{\h^2\pi}{N_c\mathcal{V}T}\!\!\sum_{\bm q, ss\p}\!\!\frac{(\omega(\bm q)_{s} {-} \omega(\bm q)_{s\p})^2}{4} v^\a\!(\bm q)_{ss^\prime} v^\b\!(\bm q)_{s\p\! s}\!\!\int\!\dd{\w}b(\bm q,\w)_{s\p}b(\bm q,-\w)_{s}  \pdv{n_T(\w)}{\hbar\w}  \:.
		\end{split}
	\end{equation}
The first and second lines evidently coincide respectively with the resonant and antiresonant thermal conductivity obtained from Hardy heat flux, introduced in \eqref{k res}-\eqref{k ares}. The procedure for the Wigner heat flux current-current response function is exactly the same as for the resonant Hardy heat flux, hence one can repeat all the steps outlined so far using \eqref{J wig app} and verify that the only difference in the derivation consists in the definition of the velocity matrix. 
Therefore, the expression for thermal conductivity obtained from Wigner definition of the heat-flux operator reads
	\begin{equation} \label{k wig app}
		\kappa^{\a\b} = \frac{\h^2\pi}{N_c\mathcal{V}T}\!\!\sum_{\bm q, ss\p}\!\!\frac{(\omega(\bm q)_{s}{+}\omega(\bm q)_{s\p})^2}{4}\tenscomp{v}^\a(\bm q)_{ss^\prime} \tenscomp{v}^\b(\bm q)_{s\p\! s}\!\!\int\!\dd{\w}b(\bm q,\w)_{s\p}b(\bm q,\w)_{s} \left[ -\pdv{n_T(\w)}{\h\w} \right],
	\end{equation}
which is exactly \eqref{k wbte}. As a side note, let us discuss the spectral integrals featuring in \eqref{k hardy app},\eqref{k wig app}. In fact, one could question the definiteness of such integrals arguing that the derivative of the Bose function diverges as ${\sim} \w^{-2}$ for $\w {\to} 0$. However, it can be shown  \cite{mahan2013many} that the bosonic spectral density must go to zero at least linearly for $\w{\to} 0$, thus ensuring the convergence of the aforementioned integrals.

\section{Details on {the} comparison {between different} theoretical description of lattice thermal transport }
\label{app:greens functions}
In this Appendix, we provide a comparison between the present {Green-Kubo approach} to derive {the expression for the} thermal conductivity $\kappa$, and other derivations proposed in the literature.
The thermal conductivity can be computed from the Kubo formula directly from its {expression}in real-time domain \eqref{kubo lambda}, that we recall below
\begin{equation}
	\label{kubo lambda app bis}
	\kappa^{\alpha\beta} = \frac{N_c\mathcal{V}}{ T} \lim\limits_{\epsilon{\to}0} \int\limits_0^\infty\!\!\dd{t} e^{{-}\epsilon t}\!\int\limits_0^\beta\!\!\dd{\lambda} \langle\hat{J}^{\beta}(0)\hat{J}^{\alpha}(t+i\hbar\lambda)\rangle\:.
\end{equation}
In order to compute explicitly the conductivity,  the so-called decoupling scheme
 \cite{isaeva2019modeling,pereverzev2018theoretical,semwal1972thermal,srivastava2019physics,pathak1965theory} is employed,  applied to the two-particle correlation function in real time domain, namely 
\begin{equation} \label{decoupling}
	\langle abcd\rangle \simeq \langle ab \rangle\langle cd \rangle+ \langle ac \rangle\langle bd \rangle + \langle ad \rangle\langle bc \rangle,    
\end{equation}
where $a,b,c,d$ are  $a(\bm q)/a^\dagger(\bm q)$ operators. 
As we shall see here, employing the decoupling scheme \eqref{decoupling} is formally equivalent to computing the response function in the dressed-bubble approximation \eqref{dressed bubble}, and will lead to the same expression of thermal conductivity in terms of the spectral densities \eqref{k hardy}-\eqref{k wbte}. 
To this aim, let's consider the case of the Hardy heat-flux operator \eqref{J hardy}. For the sake of brevity, we will just perform the calculation of the resonant thermal conductivity, hence considering only the resonant part of Hardy heat-flux operator \eqref{J res}.  Plugging the latter in \eqref{kubo lambda app bis} yields 
	\begin{equation}
		\kappa^{\alpha\beta}_R {=} \frac{\hbar^2}{N_c\mathcal{V} T} \int\limits_0^\infty\!\!\dd{t} e^{{-}\epsilon t}\!\int\limits_0^\beta\!\!\dd{\lambda}   \!\!\sum_{\bm q, ss\p}\!\sum_{\bm k, ll\p}\!\frac{\omega(\bm q)_{s} \!+\! \omega(\bm q)_{s\p}}{2}\frac{\omega(\bm k)_{l} \!+\! \omega(\bm k)_{l\p}}{2}v^\b(\bm k)_{ll^\prime} v^\a(\bm q)_{ss^\prime} \langle\hat a^\dagger\!(\bm k)_l \hat a(\bm k)_{l^\prime} \hat a^\dagger\!(\bm q,\bar{t})_s \hat a(\bm q,\bar{t})_{s^\prime} \rangle,
	\end{equation}
where for brevity we have reported $\bar{t}{=} t{+}i\hbar\lambda$ and omitted the $\epsilon{\to}0$ limit. Now we can use the decoupling scheme \eqref{decoupling} in the latter expression, obtaining (the decoupling scheme \eqref{decoupling} carries the same combination of delta functions and coefficients as in the second line of Eq.\ \eqref{dressed bubble})
	\begin{equation} \label{kappa = c c app}
		\kappa^{\alpha\beta}_R {=} \frac{\hbar^2}{N_c\mathcal{V} T}\!\!\sum_{\bm q, ss\p}\!\!\frac{(\omega(\bm q)_{s} \!+\! \omega(\bm q)_{s\p})^2}{4}v^\a(\bm q)_{ss^\prime} v^\b(\bm q)_{s\p\! s}\int\limits_0^\infty\!\!\dd{t} e^{{-}\epsilon t}\!\int\limits_0^\beta\!\!\dd{\lambda}   c(\bm q,\bar{t})_{s\p}\tilde c(\bm q,\bar{t})_{s},
	\end{equation}
where we have introduced the correlation functions $c(\bm q,{t})_{s} {=} \langle{\hat a(\bm q,{t})_s \hat a^\dagger\!(\bm q)_{s}}\rangle$ and $\tilde c(\bm q,{t})_{s} {=} \langle{\hat a^\dagger\!(\bm q,{t})_s \hat a(\bm q)_{s}}\rangle$. At this point, we note in passing that the time-domain procedure can be quite useful to understand the interband conduction as an interference mechanism. To do so, let's take Eq.\ (D4) and let's focus only on the time-integration term. Let's consider for simplicity only two different non-degenerate modes $s_1$ and $s_2$, and neglect momentum dependence and quantum correlation (the integral $\int_{0}^{\beta}\dd{\lambda}$ just yields a multiplicative $\beta$ factor). The (classical) interband thermal conductivity is then proportional to $\kappa_{R}{\propto}\int_0^\infty\dd{t}c_{s_1}(t)\tilde c_{s_2}(t)  $. If we approximate the time-dependence of the correlation functions (for $t{>}0$) as the one of the damped harmonic modes $c_{s_1}(t){\propto} e^{-\gamma_{s_1}t - i\omega_{s_1}t}$, $\tilde c_{s_2}(t){\propto} e^{-\gamma_{s_2}t + i\omega_{s_2}t}$, we get that the interband conductivity is $\kappa_{R}{\propto}\int_0^\infty\dd{t}\cos[(\omega_{s_1}-\omega_{s_2})t] e^{-(\gamma_{s_1}+\gamma_{s_2})t} $. From the latter expression, the interband conduction can be understood as interference between two different phonon modes, which of course follows from the wave-like nature of atomic oscillations. In fact, in molecular dynamics simulations, the interband contribution to thermal conductivity is deduced from the oscillatory behaviour of the heat-flux autocorrelation function \cite{pereverzev2018theoretical}. This calculation is clearly qualitative, and only meets the purpose of a more clear physical interpretation.

In order to account rigorously for the effects of interactions, the real-time averages $c(\bm q,{t})_{s},\tilde c(\bm q,{t})_{s}$ must be linked to the interacting phonon spectral densities \eqref{spectral densities}. This can be done by means of the fluctuation-dissipation theorem \cite{kubo1966fluctuation}. Indeed one can show that
\begin{subequations}
	\begin{eqnarray}
		c(\bm q,{t})_{s}= &&{-}\!\!\int\!\!\dd{\w}e^{-i\w t}\left[n_T(\w)+1\right]b(\bm q,\w)_s \label{correlation >}, \\
		\tilde c(\bm q,{t})_{s}=&& {-}\!\!\int\!\!\dd{\w}e^{i\w t}n_T(\w)b(\bm q,\w)_s \:  \label{correlation <}.
	\end{eqnarray}
\end{subequations}
We can then plug the latter relations into \eqref{kappa = c c app}, and with straightforward integration of the phase factors of the correlation functions we get
	\begin{equation}\label{k = bb app}
		\kappa^{\alpha\beta}_R {=} \frac{\hbar^2\pi}{N_c\mathcal{V} T}\!\!\sum_{\bm q, ss\p}\!\!\frac{(\omega(\bm q)_{s} \!+\! \omega(\bm q)_{s\p})^2}{4}v^\a(\bm q)_{ss^\prime} v^\b(\bm q)_{s\p\! s}\!\!\int\!\!\!\!\!\int\!\!\dd{\w_1}\!\dd{\w_2}b(\bm q,\w_1)_{s\p}b(\bm q,\w_2)_{s} \delta(\w_1{-}\w_2)\frac{n_T(\w_2)}{k_BT}\left[n_T(\w_1)+1\right],
	\end{equation}
which evidently coincides with the result from the resonant Hardy heat flux obtained in \eqref{k res} since $\frac{n_T(\w)}{k_BT}\left[n_T(\w)+1\right]{=} {-}\pdv{n_T(\w)}{\h\w}$. The calculation outlined so far proves that the decoupling scheme \eqref{decoupling} performed on real-time Green's functions coincides with the dressed-bubble approximation of the retarded response function discussed in \eqref{dressed bubble} {and} employed in imaginary-time domain. 
It follows that, at this level of approximation, adopting one scheme or another is just a matter of taste. Nonetheless, we find that the general framework based on a diagrammatic expansion in the frequency domain makes the approximation used more transparent, and can be systematically extended to include additional effects due to vertex corrections.

The approach outlined in this Appendix basically consist in the procedure proposed in Ref.\  \cite{semwal1972thermal} to approximate the Kubo formula for thermal conductivity from its time domain expression \eqref{kubo lambda app bis}. However, both in the latter work and in Ref.\  \cite{maradudin1964lattice} the Green's functions are defined using the phonon displacement operators $\hat A(\bm q)_s {=} \hat a(\bm q)_s {+} \hat a^\dagger(-\bm q)_s  $ and $\hat B(\bm q)_s {=} \hat a(\bm q)_s {-} \hat a^\dagger(-\bm q)_s$ as (in imaginary time-domain) 
\begin{subequations}
	\begin{eqnarray}
		G(\bm q,\tau)_s& = {-}\langle \mathcal{T}_\tau \hat A(\bm q,\tau)_s\hat A(-\bm q,0)_s \rangle \,, \label{G AA}\\ 
		\tilde{G}(\bm q,\tau)_s& = {-}\langle \mathcal{T}_\tau \hat B(\bm q,\tau)_s\hat A(-\bm q,0)_s \rangle \:. \label{G BA}
	\end{eqnarray}
\end{subequations}
To match the expression \eqref{k hardy}-\eqref{k ares} with the one presented in Ref.\  \cite{maradudin1964lattice}- \cite{semwal1972thermal}, one can notice the relation (valid when anomalous terms of the form $\langle \mathcal{T}_{\tau} \hat a(\tau) \hat a(0)  \rangle$ are negligible \cite{deo1966calculation,behera1967substitutional,meng2015lattice}) with the Green's functions \eqref{G AA}-\eqref{G BA} and the one we use \eqref{g(t) app}
\begin{subequations}
	\begin{eqnarray}
		G(\bm q,\tau)_s& = g(\bm q, \tau)_s +g(\bm q, -\tau)_s   \,, \label{G = g + g}\\ 
		\tilde{G}(\bm q,\tau)_s& =  g(\bm q, \tau)_s -g(\bm q, -\tau)_s   \,, \label{G = g - g}
	\end{eqnarray}
\end{subequations} 
which {imply} that the relation between the spectral densities  $B(\bm q,\omega)_s{=} {-}\frac{\h}{\pi} \Im G(\bm q,i\w_n{\to} \h\w {+}i0^+)_s $ is
\begin{subequations}
	\begin{eqnarray}
		B(\bm q,\omega)_s& = b(\bm q, \w)_s -b(\bm q, -\w)_s   \,, \label{B = b - b}\\ 
		\tilde{B}(\bm q,\omega)_s&  =   b(\bm q, \w)_s +b(\bm q, -\w)_s   \,. \label{B = b + b}
	\end{eqnarray}
\end{subequations}  
Using the latter relations to replace the spectral densities in the full thermal conductivity expression derived from Hardy heat flux \eqref{k hardy}, {and} exploiting the properties of the velocity matrices \eqref{vel properties}, {one gets}
	\begin{equation}
		\kappa^{\a\b}_{\text{\scriptsize{Ref.\  \cite{maradudin1964lattice,semwal1972thermal} }}} \!{=}  \frac{\hbar^2\pi}{N_c\mathcal{V} T}\!\!\sum_{\bm q, ss\p}\!\!v^\a(\bm q)_{ss^\prime} v^\b(\bm q)_{s\p\! s}\!\!\int\!\!\dd{\w} \left[ \w(\bm q)_s^2 B(\bm q,\omega)_sB(\bm q,\omega)_{s\p} {+}\w(\bm q)_s\w(\bm q)_{s\p} \tilde{B}(\bm q,\omega)_s\tilde{B}(\bm q,\omega)_{s\p}   \right]\!  \left[- \pdv{n_T(\w)}{\hbar\w} \right],
	\end{equation}
{which} is the expression presented in Eq.(46)-(47) of Ref.\  \cite{semwal1972thermal}  and in Eq.(4.14) of Ref.\  \cite{maradudin1964lattice}. Notice how, as anticipated in the main text, the latter expression features no clear separation between resonant and antiresonant contributions, due to the representation of the Green's functions \eqref{G = g + g}-\eqref{G = g - g} that mixes such terms.

Instead for what concerns the comparison with Ref.\  \cite{isaeva2019modeling}, we note that the authors obtained a different result starting from the same Hardy flux operator \eqref{J hardy} and implementing the Green-function method via the decoupling scheme \eqref{decoupling}. {Specifically,} the final result for conductivity presented in Ref.\  \cite{isaeva2019modeling} is
	\begin{equation}\label{k isaeva}
		\begin{split}
			&\kappa^{\a\b}_{\text{\scriptsize{Ref.\  \cite{isaeva2019modeling} }}} \!{=}  \frac{\hbar^2}{N_c\mathcal{V} T}\!\!\sum_{\bm q, ss\p}\!\!\frac{(\omega(\bm q)_{s} {+} \omega(\bm q)_{s\p})^2}{4}v^\a(\bm q)_{ss^\prime} 
			v^\b(\bm q)_{s\p\! s}\frac{n_T(\w(\bm q)_s) {-}n_T(\w(\bm q)_{s\p})}{\hbar(\w(\bm q)_{s\p}{-}\w(\bm q)_{s})}\frac{\gamma(\bm q)_s{+} \gamma(\bm q)_{s\p} \!}{ (\w(\bm q)_s{-} \w(\bm q)_{s\p}\! )^2{+} (\gamma(\bm q)_s{+} \gamma(\bm q)_{s\p}\!)^2} \\
			&\hspace{.8cm}{+}\frac{\hbar^2}{N_c\mathcal{V} T}\!\!\sum_{\bm q, ss\p}\!\!\frac{(\omega(\bm q)_{s} {-} \omega(\bm q)_{s\p})^2}{4}v^\a(\bm q)_{ss^\prime} v^\b(\bm q)_{s\p\! s}\frac{n_T(\w(\bm q)_s){-}n_T({-}\w(\bm q)_{s\p})}{\hbar(\w(\bm q)_{s\p}{+}\w(\bm q)_{s})}\frac{\gamma(\bm q)_s{+} \gamma(\bm q)_{s\p} \!}{ (\w(\bm q)_s{+} \w(\bm q)_{s\p}\! )^2{+} (\gamma(\bm q)_s{+} \gamma(\bm q)_{s\p}\!)^2},
		\end{split}
	\end{equation}
which is to be compared with the present result from Hardy heat flux in the LSFA \eqref{k hardy rta}-\eqref{k ares rta}. As briefly mentioned in the main text, the difference resides in having {derivatives of Bose-Einstein occupations with respect to frequency (which yield the modal specific heats) in Eqs.\ \eqref{k hardy rta}-\eqref{k ares rta}
	and a ``discretized'' version of such derivative (the differences of Bose-Einstein occupations divided by the  difference of the corresponding frequencies) in Eq.\ \eqref{k isaeva}.
}
The source of this discrepancy  {lies in the approximation scheme employed to 
	reduce the spectral-density-based description to a quasiparticle picture, in which phonons have well-defined energy ($\hbar\omega(\bm{q})_s$) and lifetime ($[\Gamma(\bm{q})_s]^{-1}$).}
In the present {derivation}, {we perform the dressed-bubble calculation and obtain the thermal conductivity expression \eqref{k hardy}, which relates the thermal conductivity to the spectral densities. 
	After having derived this general conductivity expression, which describes both the quasiparticle and the overdamped regime, we perform some approximations that allow one to recover a much simpler form, valid only in the regime where quasiparticles are well defined. 
	Specifically, we employ the LSFA, discarding the self-energy's real part 
	and approximating the self-energy's imaginary part with its value at the harmonic phonon frequency, so that the spectral function \eqref{spectral density and SE} becomes a Lorentzian centered at the phonon frequency $\omega(\bm{q})_s$ and with a full width at half maximum equal to the phonon linewidth $\Gamma(\bm{q})_s$. Then we approximate the derivatives of the Bose distribution with their values at the phonon frequencies where the Lorentzian spectral functions are centered. This approximation yields an expression for the thermal conductivity (Eq.\ \eqref{k wbte rta} or Eq.\ \eqref{k hardy rta}) that contains only quantities related to well-defined quasiparticles (\textit{i.e}.\ frequencies, linewidths, velocities, and specific heat).
	In contrast, Ref.\  \cite{isaeva2019modeling} employs the common approximation scheme (used e.g.\ also in Ref.\ \cite{pereverzev2018theoretical}) that enforces the validity of the quasiparticle picture with a stronger condition and earlier in the derivation, assuming the correlation functions \eqref{correlation >}-\eqref{correlation <} to have the form $c(\bm q,{t})_{s} {\simeq}[ n_T(\w(\bm q)_{s}){+}1 ]e^{i\w(\bm q)_st{-}\gamma(\bm q)_st}$ ({that is anyway} valid only for $t{>}0$). 
	Clearly, performing this approximation on the correlation functions does not allow to obtain thermal conductivity expressions 
	accounting for {the full frequency dependence of the self-energy, as it is needed} to describe transport in the overdamped regime. 
	Moreover, inserting this approximation for the correlation function into \eqref{kappa = c c app}, it is easy to see that the integral  in the inverse-temperature-like variable $\lambda$ (also called canonical correlation  \cite{kubo1966fluctuation})  produces complex phase factors that depend on $\gamma(\bm q)_s$ and must be neglected to recover the result \eqref{k isaeva}.}
From a practical viewpoint, numerical results presented in Sec.\ \ref{subsec:applications} showed that no quantitative difference {originates} from this formal discrepancy in the expressions for the interband thermal conductivity, at least in the test case materials considered.  
{From a formal viewpoint, the procedure employed in this work has some useful features that are worth to be highlighted. In particular, it allows to: (i) obtain a thermal conductivity expression accounting for spectral functions and thus apt to describe also the overdamped regime of thermal transport; (ii) discuss in detail all the approximations required to extract from the spectral function a well-defined quasiparticle lifetime, thus to obtain a thermal conductivity expression including only quasiparticles' properties (frequencies, linewidths, velocities, specific heats). 
}

In conclusion, it is worth mentioning that in Ref.\  \cite{pereverzev2018theoretical}, a Green-Kubo approach was used to derive {a} thermal conductivity {expression} employing yet another {original formula for the} heat flux. 
{Specifically, the heat flux expression employed in Ref.\  \cite{pereverzev2018theoretical} } stems from a modification of Hardy on-site energy \eqref{hardy h(R)}, and has an overall structure analogous to Hardy heat flux \eqref{J hardy}:
\begin{subequations}
	\begin{eqnarray}
		&&\bm{\hat{J}_{\text{\small Ref.\  \cite{pereverzev2018theoretical}}}} \pp= \bm{\hat{J}}_R\pp + \bm{\hat{J}}_A\pp  \:,\label{J pereverzev} \\
		\label{J res pereverzev}
		&&\begin{split}
			\bm{\hat{J}}_R\pp \! =\bm{\hat{J}}_R {+}\!\frac{\h}{N_c\mathcal{V}}\!\!\sum_{\bm q,ss\p}\!\!\frac{\omega(\bm q)_{s} {-} \omega(\bm q)_{s\p}}{2}  & \bm v\pp\!(\bm q)_{ss^\prime} \hat a^\dagger\!(\bm q)_s \hat a(\bm q)_{s^\prime}\, ,
		\end{split} \\
		\label{J ares pereverzev}
		&&\begin{split}
			\bm{\hat{J}}_A\pp\! =\bm{\hat{J}}_A{+} \!\frac{\h}{N_c\mathcal{V}}\!\sum_{\bm q,ss\p}\!& \frac{\omega(\bm q)_{s}\! {+} \omega(\bm q)_{s\p}}{4} \bm v\pp\!(\bm q)_{ss^\prime} \left[\hat a(-\bm q)_s \hat a(\bm q)_{s^\prime}\!-\!\hat a^\dagger(\bm q)_s\hat a^\dagger(-\bm q)_{s^\prime} \right]\!,
		\end{split}
	\end{eqnarray}
\end{subequations}
where the extra velocity matrix is defined as
\begin{equation}
	\label{velocity pereverzev}
	\begin{split}
		\bm v\pp\!(\bm q)_{ss^\prime}{=}\frac{-i}{2\sqrt{\w(\bm q)_s\w(\bm q)_{s\p}}}\sum\limits_{\bm R}\frac{\bm \Phi_{\bm Rb\a,0b\p\!\a\p}}{M_b}\!(\bm R{+}\bm{\tau}_b{-}\bm{\tau}_{b\p})\mathcal{E}^*\!(\bm q)_{s,b\a}\mathcal{E}(\bm q)_{s\p\!,b\p\!\a\p}   \:.
	\end{split}
\end{equation}

We did not perform the test to compare the result for interband conductivity proposed by Ref.\  \cite{pereverzev2018theoretical}. However, we expect small correction from the extra terms in the heat flux \eqref{J pereverzev}.{ In fact, we have discussed how the resonant (antiresonant) current-current response function is peaked when $\w(\bm q)_s{-}\w(\bm q)_{s\p}{\simeq}0$ ($\w(\bm q)_s{+}\w(\bm q)_{s\p}{\simeq}0$), but the extra term in the resonant (antiresonant) heat flux \eqref{J res pereverzev}(\eqref{J ares pereverzev}) is proportional to $\w(\bm q)_s{-}\w(\bm q)_{s\p}$ ($\w(\bm q)_s{+}\w(\bm q)_{s\p}$), hence all the contributions from it to the thermal conductivity {are expected to} be small for the relevant coupling modes.} Notice also that if the correction to the velocity matrix \eqref{velocity pereverzev} is neglected, the final result for thermal conductivity presented in Ref.\  \cite{pereverzev2018theoretical} coincides with the one presented in Ref.\  \cite{isaeva2019modeling} and briefly discussed here in \eqref{k isaeva}.
\twocolumngrid

\bibliography{kubobib}

\end{document}